\documentclass[aps,prd,preprint,groupedaddress]{revtex4}

\usepackage{graphicx}
\usepackage{amsmath}
\usepackage{bm}

\def\ee{\end{eqnarray}}

\def\=:{=\hspace{-.7em}\raisebox{1.1ex}{.}\hspace{.1em}\raisebox{-0.2ex}{.} }

\newcommand {\beq}{\begin{eqnarray}}
\newcommand {\eeq}{\end{eqnarray}}
\newcommand {\non}{\nonumber\\}

\newcommand {\1}[1]{\frac{1}{#1}}

\newcommand {\ph}{\varphi}

\newcommand {\del}{\partial}

\newcommand {\tr}{{\rm tr}\,}

\newcommand{\hs}[1]{\hspace{#1 mm}}

\begin{document}


\title{
Knotted instantons\\ 
from annihilations of monopole-instanton complex 
}


\author{Muneto Nitta$^{1}$}

\affiliation{
$^1$Department of Physics, and Research and Education Center for Natural 
Sciences, Keio University, Hiyoshi 4-1-1, Yokohama, Kanagawa 223-8521, Japan\\
}


\date{\today}
\begin{abstract}
Monopoles and instantons are sheets (membranes) and strings in $d=5+1$, 
respectively, and instanton strings can terminate on monopole sheets.
We consider a pair of monopole and anti-monopole sheets 
which is unstable to decay and results in 
a creation of closed instanton strings.
We show that when an instanton string is stretched between 
the monopole sheets, 
there remains a new topological soliton of codimension five 
after the pair annihilation, 
{\it i.e.}, a twisted closed instanton string or a knotted instanton.

\end{abstract}

\maketitle

\section{Introduction}

Yang-Mills instantons and monopoles are one of the 
important topological solitons in gauge field theories 
since they play central roles in non-perturbative dynamics 
of supersymmetric gauge theories.  
't Hooft-Polyakov monopoles are 
point-like topological defects in $SU(2)$ Yang-Mills theory, 
coupled with Higgs scalar fields 
in the triplet representation in 
$d=3+1$ dimensions \cite{'tHooft:1974qc,Polyakov:1974ek}.
In particular, Bogomol'nyi-Prasad-Sommerfield (BPS) monopoles 
appear as classically and quantum mechanically stable 
solitons in supersymmetry gauge theories.
Yang-Mills instantons are point-like 
in Euclidean four dimensional pure Yang-Mills theory 
\cite{Belavin:1975fg}.
Instantons and monopoles are 
point particles and strings in $d=4+1$, respectively, 
while they are 
strings and sheets (or membranes) in $d=5+1$, respectively. 
Instantons and BPS monopoles are closely related to each other.
First, BPS monopole equations are given by dimensional reduction 
of (anti-)self-dual equations for instantons along one direction.
When one direction of space is compactified to a circle, 
caloron solutions, {\it i.e.}, periodic instantons, 
interpolate between instantons and BPS monopoles. 
In string theory,
instantons regarded as D0-branes inside D4-branes in type II string theory \cite{Witten:1995gx,Douglas:1995bn} and 
BPS monopoles regarded as D1-branes stretched between D3-branes 
in type II string theory \cite{Green:1996qg,Diaconescu:1996rk} 
are related by a T-duality.

Because of the above mentioned relation between instantons and BPS monopoles, 
a BPS monopole can be understood to be made of infinite number of instantons. 
To see this relation more clearly, 
let us consider a configuration of a monopole string and an anti-monopole string  placed on a parallel with each other in $d=4+1$ dimensions.
This configuration is unstable to decay.
While a decay of metastable solitons was studied before in field theory 
\cite{Preskill:1992ck,Vilenkin:1982hm}, a pair annihilation 
of a soliton and an anti-soliton has not been studied very well.
As anhihilations of a D$p$-brane and an anti-D$p$-brane in collision 
result in a creation of D$(p-2)$-branes \cite{Sen:2004nf}, 
a collision of monopole stings  
is not necessary to reach complete annihilation. 
Rather, there can remain (anti-)instantons in general \cite{Nitta:2012kj}.
The production probability of instantons is maximized 
when the monopole strings have opposite $U(1)$ moduli \cite{Takeuchi:2012ee,Takeuchi:2012ec}. 
This was shown by putting the system into the Higgs phase 
with introducing the Fayet-Iliopoulos parameter. 
In the Higgs phase, instantons or monopoles cannot exist alone.
Instead, they can live inside a non-Abelian vortex 
\cite{Hanany:2003hp,Auzzi:2003fs,Eto:2005yh,Eto:2006cx}\footnote{
For a review of non-Abelian vortices, see \cite{Tong:2005un,Eto:2006pg,Shifman:2007ce,Shifman:2009zz}.
} 
inside which instantons and monopoles can be regarded  \cite{Eto:2004rz,Fujimori:2008ee,Tong:2003pz} as sigma model instantons (lumps) \cite{Polyakov:1975yp} 
and kinks \cite{Abraham:1992vb,Abraham:1992qv,Arai:2002xa,Arai:2003es}, 
respectively.
By doing this, the problem was reduced to 
annihilations of a domain wall and an anti-domain wall 
resulting in a creation of lumps inside a non-Abelian vortex.

In this paper, we consider 
a collision of a monopole sheet and an anti-monopole sheet in $d=5+1$, 
which results in a creation of closed instanton strings. 
Instanton strings can terminate on or stretch monopole sheets in $d=5+1$. 
We construct solutions of monopole and anti-monopole sheets stretched 
by instanton strings 
(and attached by instanton string from outside).  
The idea to show this is putting the system into the Higgs phase 
by turning on the Fayet-Iliopoulos parameter. 
Then, a non-Abelian vortex can appear, and 
an instanton string and a monopole sheet can be realized as a lump string and a domain wall 
in the non-Abelian vortex, respectively, as mentioned above. 
We find that when the monopole sheets annihilate in collision, 
there remains a new topological object 
of codimension five, that is, 
a ``twisted" closed instanton string.
Since it carries the Hopf charge of $\pi_3(S^2)$
as a knot soliton or a Hopfion does \cite{Faddeev:1996zj,Radu:2008pp} 
in the vortex world-volume,  
this new solution can be called a ``knotted instanton."
The same mechanism works in the different models; 
Annihilations of 
a pair of a domain wall and an anti-domain wall result in  
creations of knot solitons (Hopfion) \cite{Nitta:2012kk} 
in the Faddeev-Skyrme model \cite{Faddeev:1996zj}
and vortons \cite{Davis:1988jq,Radu:2008pp} 
in two component Bose-Einstein condensates \cite{Nitta:2012hy}.

This paper is organized as follows.
In Sec.~\ref{sec:instanton-monopole} we present 
composite states of monopoles and instantons 
inside a non-Abelian vortex in the Higgs phase.
In the first subsection, we construct 
monopoles, a monopole-anti-monopole pair, and instantons 
in $d=4+1$ and $5+1$ dimensions.
In the second subsection, we construct 
instanton strings terminating on monopole sheets,  
and a pair of monopole and anti-monopole sheets 
stretched by instanton strings in $d=5+1$ dimensions. 
In Sec.~\ref{sec:pair-annihilations}, 
we study pair annihilations of a monopole and an anti-monopole. 
In the first subsection, we find that (anti-)instantons 
are created in pair annihilations of 
monopole and anti-monopole strings in $d=4+1$, 
and closed instanton strings are created  
in pair annihilations of 
monopole and anti-monopole sheets 
without stretched instanton strings in $d=5+1$. 
In the second subsection, we find that 
a knotted instanton is created 
when instanton strings are stretched between 
the monopole and anti-monopole sheets. 
Section \ref{sec:summary} is devoted to summary 
and discussion.

\newpage
\section{Instanton strings and monopole sheets \label{sec:instanton-monopole}}
\subsection{Non-Abelian vortices}
We consider 't Hooft-Polyakov monopole strings (sheets) 
in the BPS limit in an $SU(2)$ gauge theory 
in $d=4+1$ ($5+1$) dimensions.
We put the system into the Higgs phase 
by considering the $U(2)$ gauge theory instead of $SU(2)$, 
where magnetic fluxes from a monopole are squeezed 
into vortices \cite{Tong:2003pz}, 
which are membranes (3-branes) in $d=4+1$ ($5+1$) dimensions.
The Lagrangian in $d=5+1$ ($4+1$ or $3+1$) dimensions, 
which we consider, is given by ($A,B=0,1,\cdots,d$)
\beq
&& {\cal L} = - \1{4 g^2}\tr F_{AB}F^{AB} 
 + \1{2g^2} \tr (D_{A} \Sigma)^2 
  + \tr D_{A}H^\dagger D^{A}H - V ,\non
&& V = g^2 \tr (H H^\dagger -v^2{\bf 1}_2)^2 
 + \tr \left[H(\Sigma - M)^2 H^\dagger\right],
 \quad  \label{eq:Lagrangian}
\eeq
with two complex scalar fields in the 
fundamental representation of $SU(2)$,  
summarized as a two by two complex matrix $H$, 
and with the real matrix scalar field $\Sigma$
in the adjoint representation. 
The mass matrix is given by  
$M={\rm diag.}(m_1,m_2)$, with $m_1>m_2$ and $m_1-m_2\equiv m$.
The constant $v^2$ is called the Fayet-Iliopoulos parameter 
in the context of supersymmety.
In the limit of vanishing $v^2$, 
the system goes back into the unbroken phase 
and $H$ decouples in the vacuum.
In $d=3+1,4+1$ dimensions, this Lagrangian 
can be made ${\cal N}=2$ supersymmetric 
({\it i.e.}, with eight supercharges)
by suitably adding fermions, while 
in $d=5+1$ dimensions, 
it can be made ${\cal N}=2$ supersymmetric 
only in the massless case $m=0$;  
however, supersymmetry is not essential in our study.

In the massless limit $m=0$, 
the Lagrangian (\ref{eq:Lagrangian}) 
enjoys the $SU(2)_{\rm F}$ flavor symmetry which acts on 
$H$ from the right side. 
In this case, the system is in so-called the color-flavor locked 
vacuum, in which both the color $U(2)$ and flavor $SU(2)_{\rm F}$ 
symmetries are spontaneously broken, 
with the color-flavor locked symmetry 
$SU(2)_{\rm C+F}$ remaining. 
In this case, the model admits a non-Abelian $U(2)$ vortex solution 
$H = {\rm diag.}\, (f(r) e^{i\theta}, v)$, 
where $(r,\theta)$ are polar coordinates \cite{Hanany:2003hp,Auzzi:2003fs}.
The vortex solution breaks the vacuum symmetry 
$SU(2)_{\rm C+F}$ into $U(1)$ in the vicinity of the vortex, 
and there appear ${\bf C}P^1 \simeq SU(2)_{\rm C+F}/U(1)$ 
Nambu-Goldstone modes localized on the vortex.
The low-energy effective theory on 
the $d-2$-dimensional vortex world-volume is the ${\bf C}P^1$ model. 
In the presence of mass, {\it i.e.} $m\neq 0$, 
the $SU(2)_{\rm C+F}$ symmetry is explicitly broken, 
inducing the same type of mass matrix in the $d-2$ dimensional 
vortex effective theory \cite{Hanany:2003hp,Auzzi:2003fs,Tong:2003pz,Eto:2004rz,Eto:2006uw}  
($\mu=0,1,\cdots,d-2$)
\beq
&& {\cal L}_{\rm vort.eff.}
=  2 \pi v^2 |\del_{\mu} z_0|^2 + {4\pi \over g^2} \left[ 
 {\partial_{\mu} u^* \partial^{\mu} u - m^2 |u|^2 
  \over (1 + |u|^2)^2} \right] .\label{eq:vortex-th}
\eeq
Here $z_0(x^{\mu})$ corresponds to the position moduli, and 
the projective coordinate $u(x^{\mu})$ of ${\bf C}P^1$ 
corresponds to the orientational moduli.
The vacua in the vortex theory are $u=0$ and $u=\infty$ corresponding to 
the north and south poles of the target space 
$S^2 \simeq {\bf C}P^1$.

This is known as the massive ${\bf C}P^1$ model, 
which can be made supersymmetric by adding fermions 
\cite{Abraham:1992vb,Abraham:1992qv,Arai:2002xa,Arai:2003es}, 
because the vortex is BPS and  
preserves/breaks a half of the original supersymmetry 
when the original theory is made 
${\cal N}=2$ supersymmetric.
However, supersymmetry is not essential in our study. 

\subsection{Monopoles and instantons inside a non-Abelian vortex}
\begin{figure}[h]
\begin{center}
\begin{tabular}{ccc}
\includegraphics[width=0.4\linewidth,keepaspectratio]{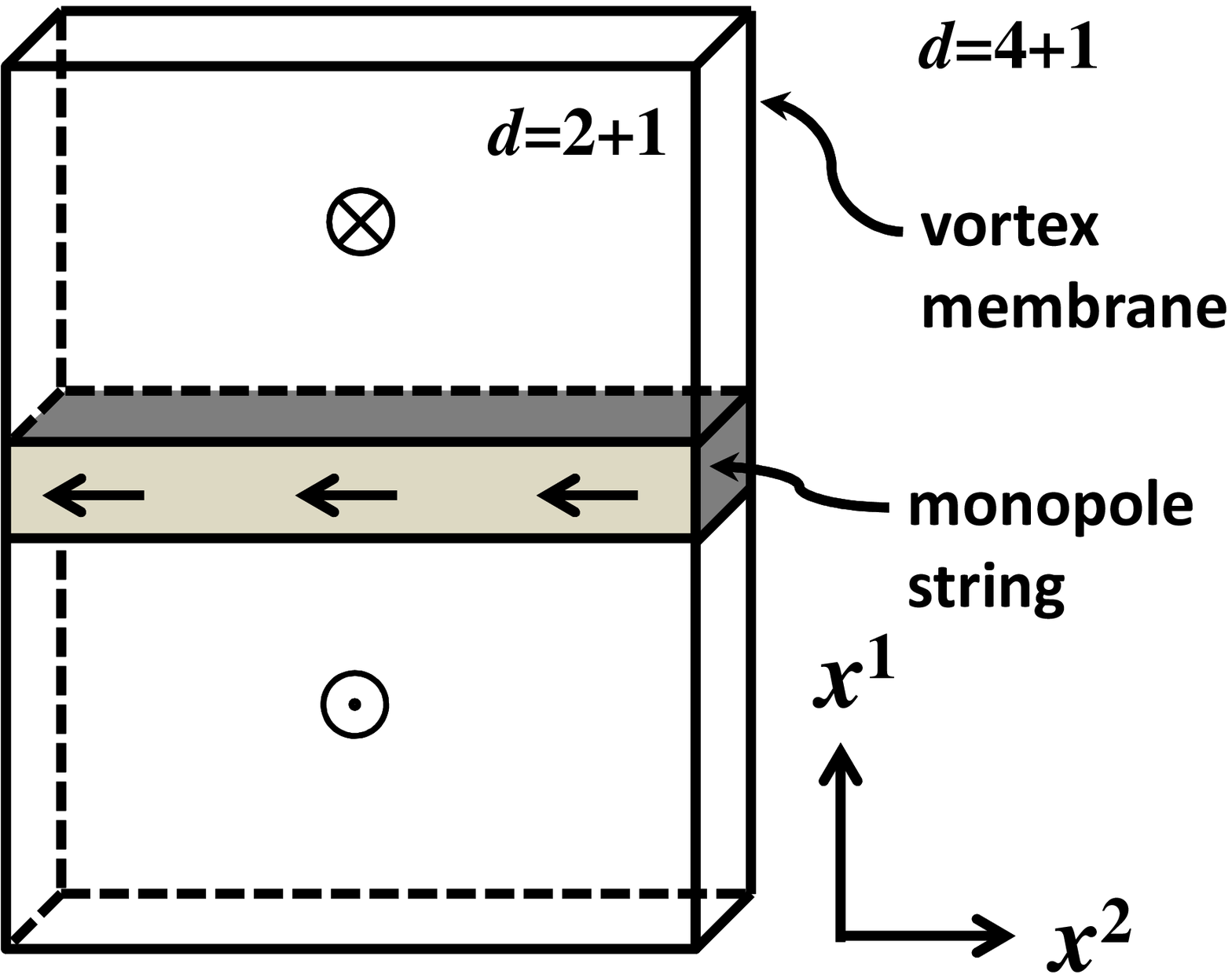}
&
\includegraphics[width=0.38\linewidth,keepaspectratio]{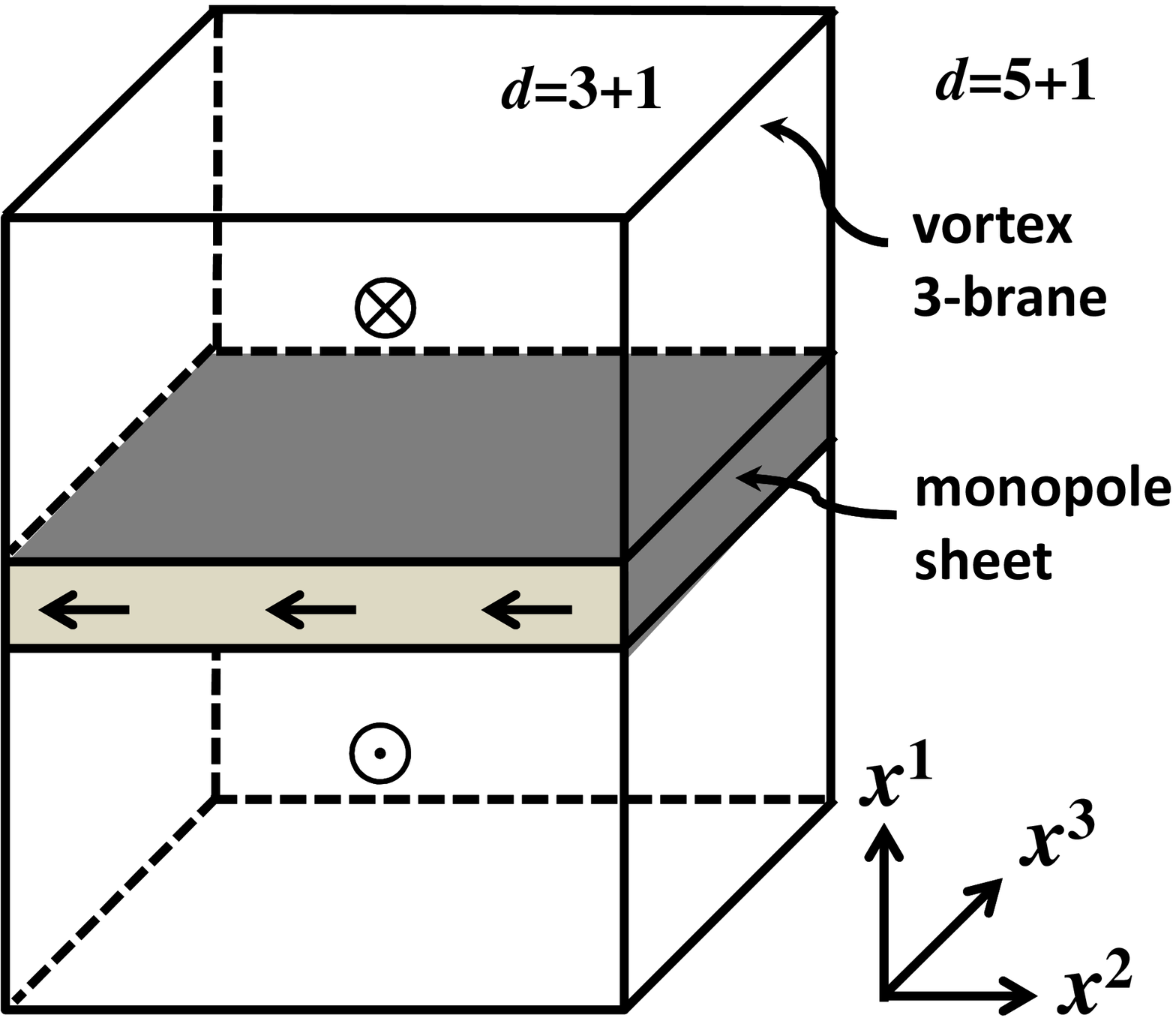}&
\includegraphics[width=0.2\linewidth,keepaspectratio]{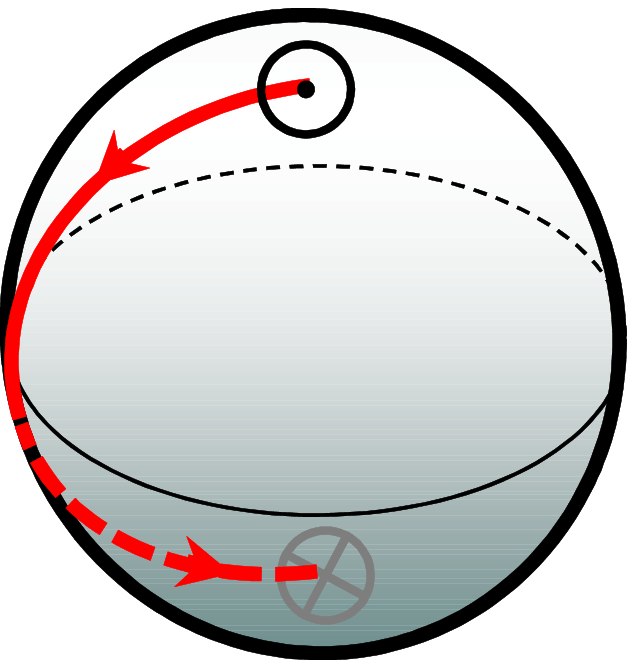}\\
(a) & (b) & (c)
\end{tabular}
\caption{(a) The monopole string in $d=4+1$. 
(b) The monopole sheet in $d=5+1$. (c) The ${\bf C}P^1$ target space of the vortex world sheet(volume). 
(a,b) 
The monopole string(sheet) is perpendicular to the $x^1$-axis.
The arrows denote points in the ${\bf C}P^1$.
(c)
The north and south poles are denoted by 
$\odot$ and $\otimes$, respectively.
The path connecting them represents the map from 
the path in (a) along the $x^1$-axis 
in real space from $x^1 \to - \infty$ to $x^1 \to + \infty$. 
The path in the ${\bf C}P^1$ target space 
passes through one point on the equator, 
which is represented by $\leftarrow$ in (a) in this example. 
In general, the $U(1)$ zero mode is localized 
on the monopole as a phase modulus.
}
\label{fig:brane-2d} 
\end{center}
\end{figure}

Let us construct a monopole solution as a domain wall in the effective theory 
(\ref{eq:vortex-th}) on the vortex.
Then, a domain wall interpolating the two vacua 
$u=0$ and $u=\infty$ can be obtained as \cite{Abraham:1992vb,Abraham:1992qv,Arai:2002xa,Arai:2003es}
\beq
 u_{\rm m}(x^1) = e^{\mp m (x^1-x^1_0) + i \ph}  \label{eq:wall-sol}
\eeq
where $\mp$ represents a wall and an anti-wall 
with the width $1/m$.
Here, $x^1_0$ and $\ph$ are real constants 
representing the position and phase [the $U(1)$ modulus] 
of the (anti-)monopole, see Fig.~\ref{fig:brane-2d}.
 The monopole configuration is mapped to 
a large circle starting from the north pole, 
denoted by $\odot$,
and ending at the south pole, 
denoted by $\otimes$, in the ${\bf C}P^1$ target space 
in Fig.~\ref{fig:brane-2d}(c).
The domain wall tension is 
\beq
 E_{\rm dw}= {4\pi \over g^2} \times m = E_{\rm m},
\eeq 
which coincides with the monopole mass $E_{\rm m}$.
It is equivalent to the monopole charge 
because the monopole is a BPS state. 
Therefore, the wall in the vortex theory 
is nothing but a monopole string (sheet) from  
the bulk point of view \cite{Tong:2003pz}. 

\begin{figure}
\begin{center}
\begin{tabular}{ccc}
\includegraphics[width=0.4\linewidth,keepaspectratio]{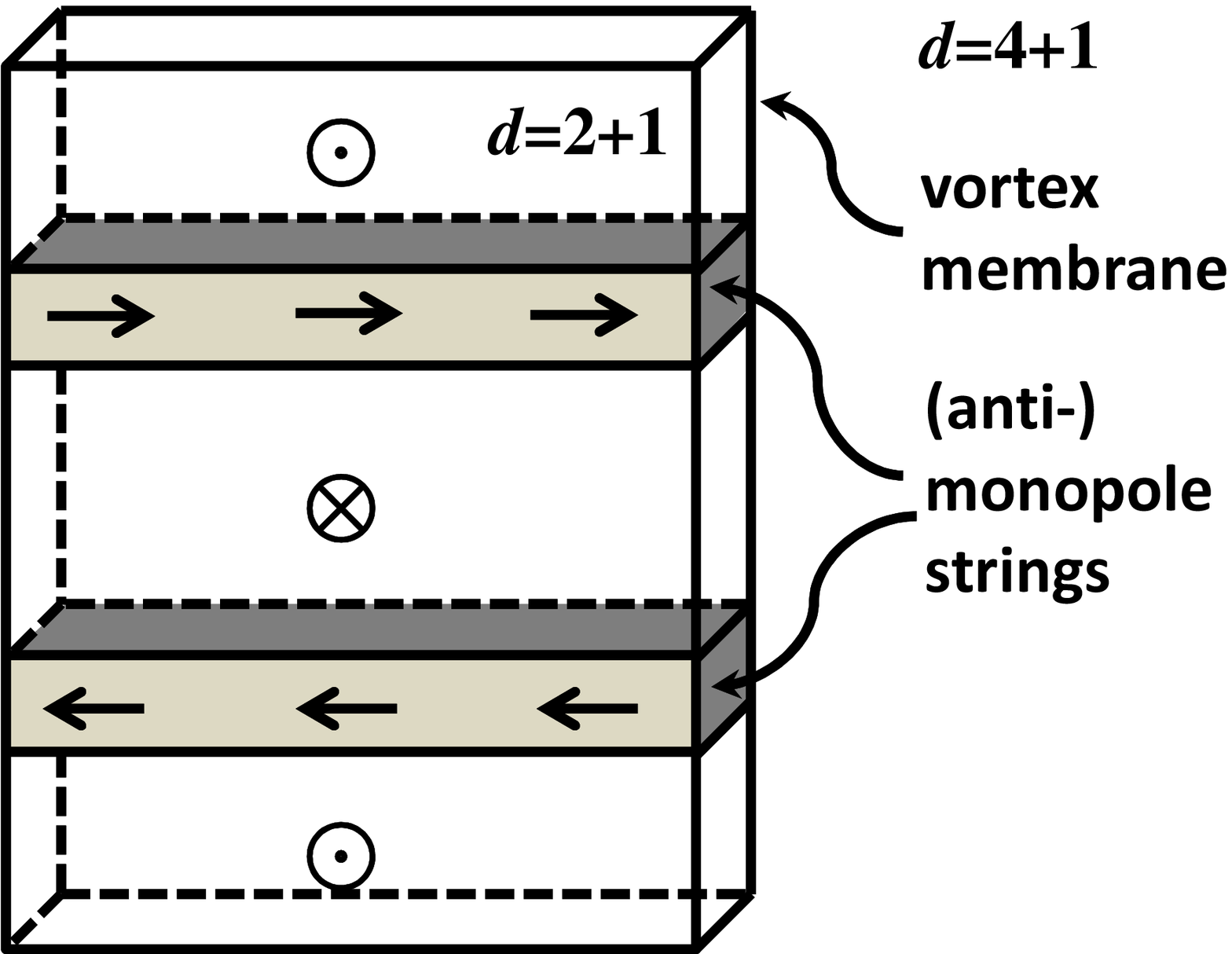}
&
\includegraphics[width=0.4\linewidth,keepaspectratio]{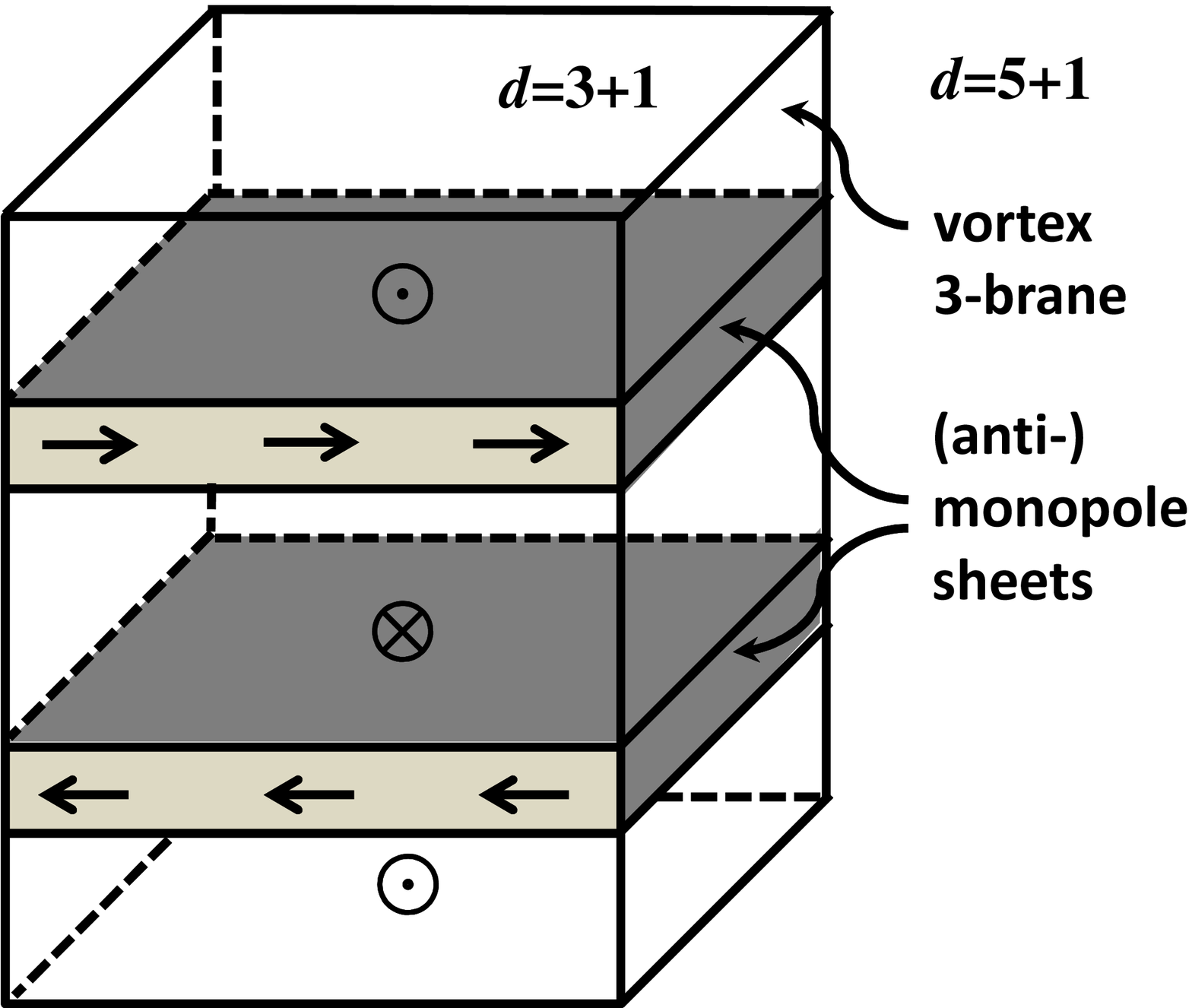}&
\includegraphics[width=0.2\linewidth,keepaspectratio]{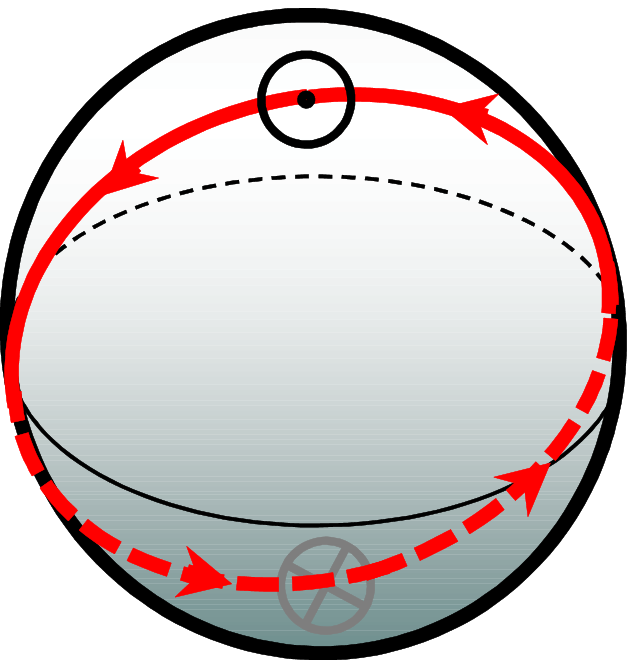}\\
(a) & (b) & (c)
\end{tabular}
\caption{
(a) Monopole and anti-monopole strings in $d=4+1$. 
(b) Monopole and anti-monopole sheets in $d=5+1$. 
(c) The ${\bf C}P^1$ target space of the vortex world sheet(volume). 
(a,b) 
The monopole and anti-monopole strings(sheets) are 
perpendicular to the $x^1$-axis.
The arrows denote points in the ${\bf C}P^1$.
(c)
The path connecting them represents the map from 
the path in (a) along the $x^1$-axis 
in real space from $x^1 \to - \infty$ to $x^1 \to + \infty$. 
The path in the ${\bf C}P^1$ target space 
passes through one point on the equator, 
which is represented by $\leftarrow$ in (a) in this example. 
In general, the $U(1)$ zero mode is localized 
on the wall as a phase modulus.
}
\label{fig:brane-anti-brane-2d} 
\end{center}
\end{figure}

Second, 
we construct a pair of 
a monopole and an anti-monopole  placed on a parallel with each other 
at $x^1=x^1_1$ and $x^1 = x^1_2$, respectively. 
An approximate solution valid at large distance, 
$x^1_2-x^1_1 \gg m^{-1}$, is obtained as 
the corresponding wall and anti-wall solution \cite{Nitta:2012hy,Nitta:2012kk}
\beq
 u_{{\rm m}-{\rm am}} (x^1)
= e^{- m (x^1-x^1_1) + i \ph_1} +  e^{+ m (x^1-x^1_2) + i \ph_2}.
\label{eq:wall-anti-wall-sol}
\eeq
Here, the phases or the $U(1)$ zero modes $\ph_1$ and $\ph_2$ 
of the monopole and anti-monopole are arbitrary. 
Later, we consider the case that they are opposite to each other,  
$\ph_1=\ph_2 +\pi$, 
as in Fig.~\ref{fig:brane-anti-brane-2d}.  
In this case, the configuration is mapped to a large circle 
in the ${\bf C}P^1$ target space, 
as in Fig.~\ref{fig:brane-anti-brane-2d}(c).  
A similar configuration was studied before on a circle 
in the $x^1$-direction, 
and the classical (in)stability was discussed \cite{Eto:2004zc}.  

On the other hand, Yang-Mills instantons in the bulk 
are sigma model lumps with 
the topological charge in $\pi_2 ({\bf C}P^1)$ \cite{Polyakov:1975yp}
in the vortex effective theory \cite{Eto:2004rz,Fujimori:2008ee}.
This can also be inferred from the lump energy, 
which coincides with the instanton energy; 
The K\"ahler class (the coefficient of the kinetic term) 
${4\pi \over g^2}$ in the vortex theory (\ref{eq:vortex-th}) 
implies that the energy $2\pi k$ of $k$ lumps inside the vortex theory 
reproduces the energy $E_{\rm i}$ of $k$ instantons in the bulk \cite{Eto:2004rz,Fujimori:2008ee}:
\beq
 E_{\rm l} = {4\pi \over g^2} \times 2\pi k 
= {8\pi^2 \over g^2} k = E_{\rm i}. 
\eeq
Since they are BPS states, this coincidence implies  
the coincidence between the topological charges $\pi_2 ({\bf C}P^1)$ and 
$\pi_3(SU(2))$ of lumps and instantons, respectively. 
With a complex coordinate $z\equiv x^2 + i x^3$, 
the instanton solutions are given as
\begin{equation}
u(z)
 = {{\prod_{j=1}^{k_+} (z - z_j^+)
\over
\prod_{i=1}^{k_-} (z - z_i^-)}}, \label{eq:lumps}
\end{equation}
with the moduli parameters $z_i^\pm$ and
the instanton number $k={\rm max}(k_+,k_-)$.

\subsection{Composite states of monopole and instantons} 
\begin{figure}
\begin{center}
\begin{tabular}{cc}
\includegraphics[width=0.3\linewidth,keepaspectratio]{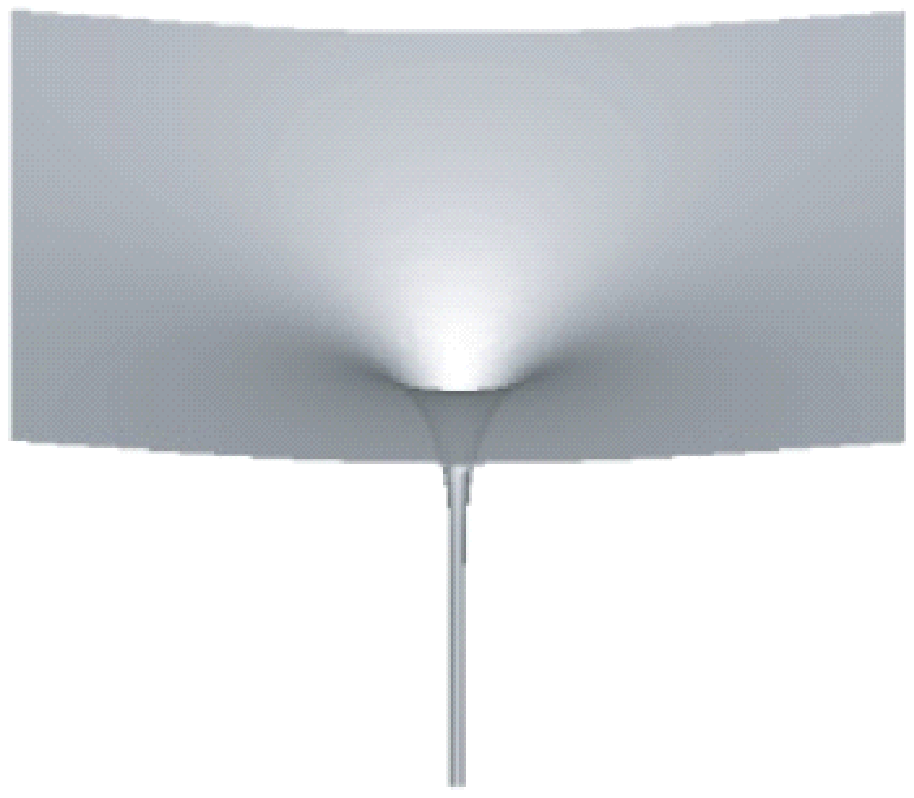} &
\includegraphics[width=0.35\linewidth,keepaspectratio]{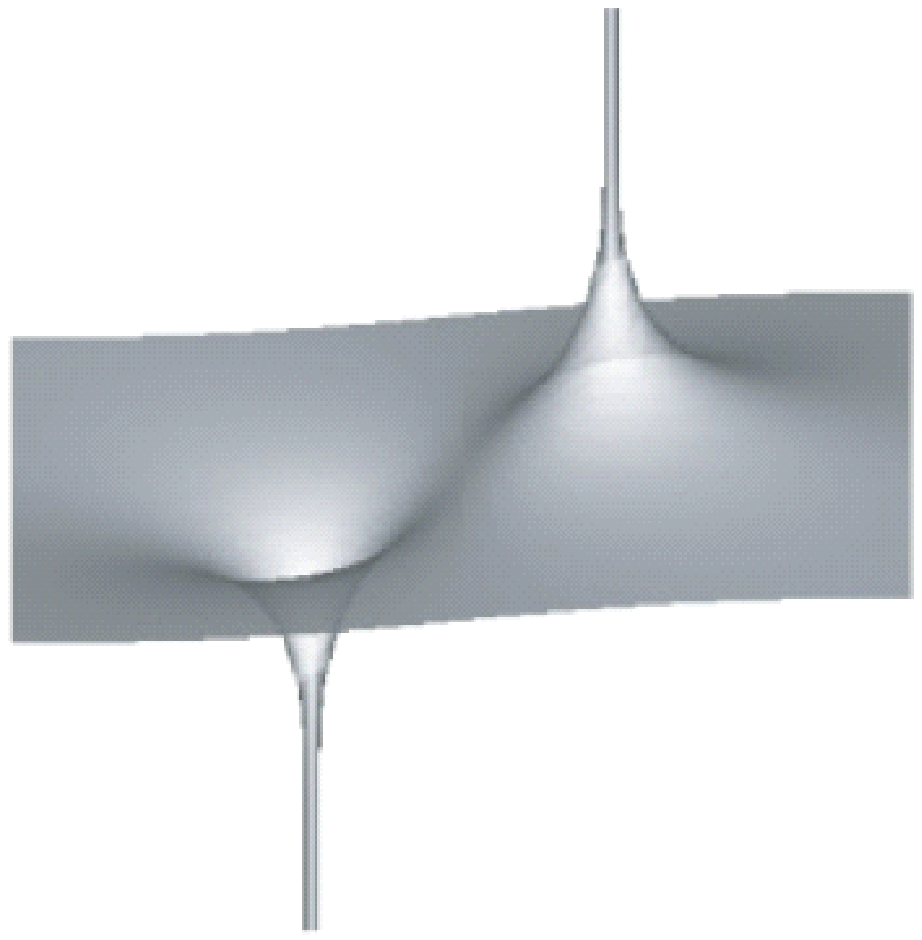} \\
(a) & (b)
\end{tabular}
\end{center}
\caption{Analytic exact solutions of instanton strings and a monopole sheet 
(the $|u|=1$ surface). 
(a) One instanton string is attached to a monopole sheet,  
given by Eq.~(\ref{eq:D-brane-soliton}) with Eq.~(\ref{eq:D-brane-soliton1}).
(b) Two instanton strings are attached to a monopole sheet 
from its both sides, given by 
Eq.~(\ref{eq:D-brane-soliton}) with Eq.~(\ref{eq:D-brane-soliton2}).
\label{fig:brane-anti-brane-with-string1} 
} 
\end{figure}

In this subsection, we construct composite states 
made of monopole sheets and instantons strings 
in $d=5+1$ dimensions. 
First, we consider $k$ instanton strings 
ending on a monopole sheet in $d=5+1$ 
as $k$ lump strings ending on a domain wall in the vortex effective theory. 
Such a configuration can be constructed as 
\cite{Gauntlett:2000de,Shifman:2002jm,Isozumi:2004vg}
\beq
u_{\rm m-i} (x^1,z)
= e^{\pm m (x^1 - x^1_0) + i \ph_0}Z(z), 
\label{eq:D-brane-soliton}
\eeq
with 
\beq
 Z(z) = \sum_{i=1}^k {\lambda_i \over z- z_i}.
\label{eq:D-brane-soliton1}
\eeq
This solution precisely coincides with 
a BIon \cite{Callan:1997kz,Gibbons:1997xz,Hashimoto:1997px} 
in the Dirac-Born-Infeld action on a D2-brane, 
and so, this was referred to as a ``D-brane soliton" 
\cite{Gauntlett:2000de}. 
The monopole sheet surface in the above solution is logarithmically bent, 
as in Fig.~\ref{fig:brane-anti-brane-with-string1}(a). 

We can also place the instanton strings on the both sides of 
the monopole sheet with Eq.~(\ref{eq:D-brane-soliton}) 
by using \cite{Isozumi:2004vg,Eto:2006pg}
\begin{equation}
Z(z)
 = {{\prod_{j=1}^{k_+} (z - z_j^+)
\over
\prod_{i=1}^{k_-} (z - z_i^-)}}, 
\label{eq:D-brane-soliton2}
\end{equation}
where $z_i^\pm$ and $k_\pm$  
denote the positions and numbers of 
instanton strings extending to $x^1 \to \pm \infty$, 
respectively. 
If the numbers of the instanton strings coincide 
on both sides, {\it i.e.}, $k_1=k_2$, 
the monopole sheet surface is asymptotically flat, 
as in Fig.~\ref{fig:brane-anti-brane-with-string1}(b).

\begin{figure}
\begin{center}
\begin{tabular}{cc}
\includegraphics[width=0.55\linewidth,keepaspectratio]{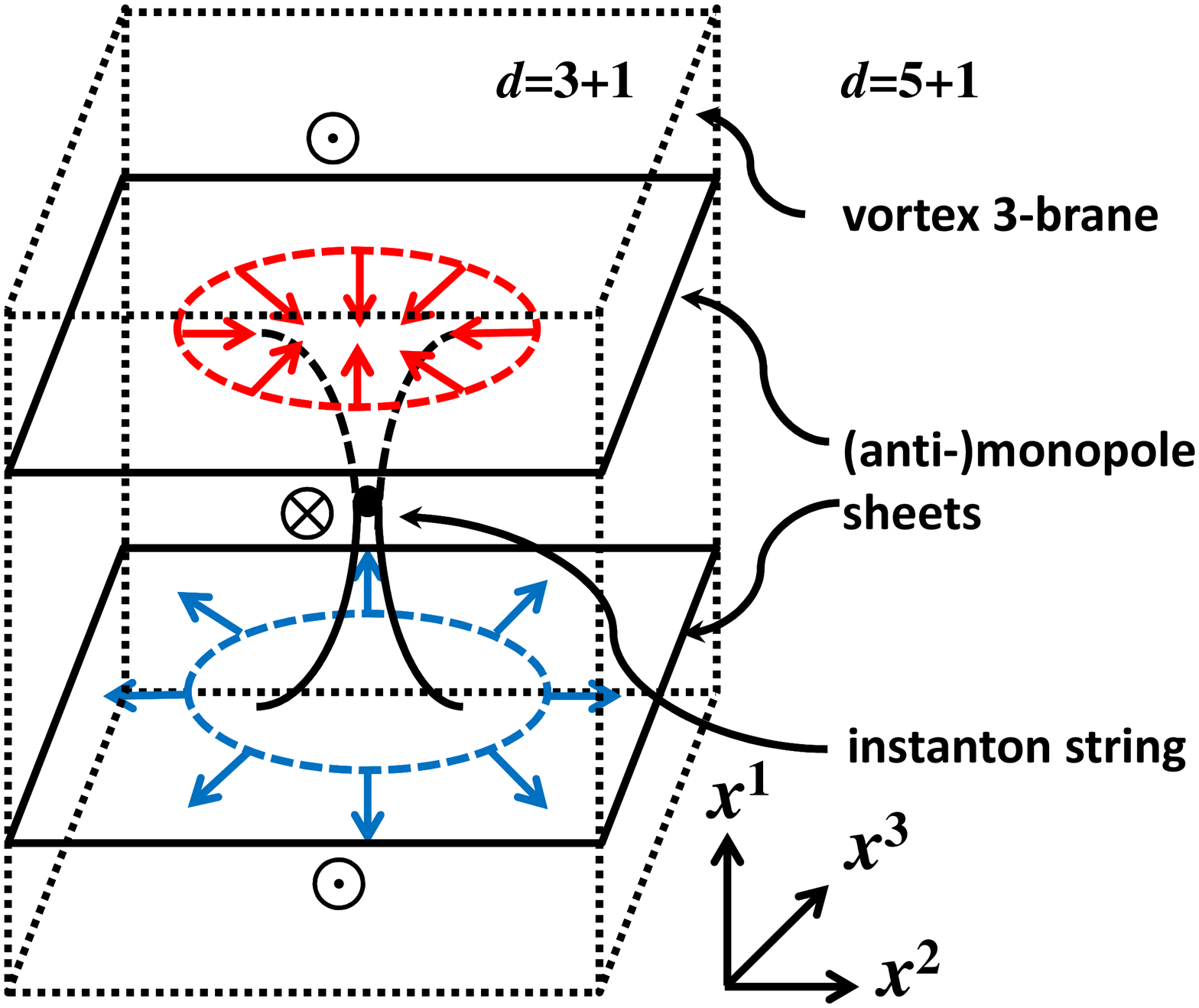} &
\includegraphics[width=0.35\linewidth,keepaspectratio]{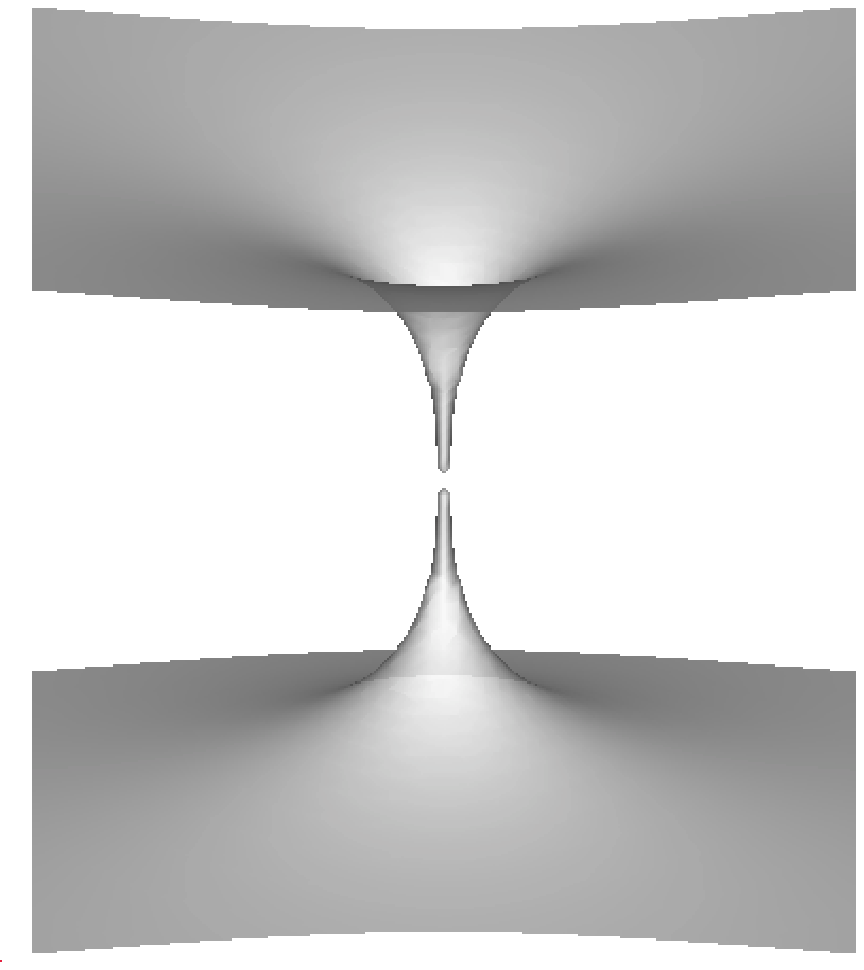}\\
(a) & (b)
\end{tabular}
\end{center}
\caption{Instanton string stretched between a pair of monopole and anti-monopole sheets given by Eq.~(\ref{eq:monopole-instanton-monopole}) 
(a) Spin structure (b) Analytic solution for well separated monopole sheets 
(the $|u|=1$ surface).
\label{fig:brane-anti-brane-with-string2} 
} 
\end{figure}

Finally let us consider the instanton strings stretched between 
the monopole sheets 
in Fig.~\ref{fig:brane-anti-brane-with-string2}. 
An approximate solution for well-separated 
monopole and anti-monopole sheets 
placed at $x^1=x^1_1$ and $x^1=x^1_2$ ($x^1_2-x^1_1 \gg m^{-1}$) 
with the phases $\ph_1$ and $\ph_2$, respectively, 
can be obtained from Eq.~(\ref{eq:wall-anti-wall-sol}) as 
\cite{Nitta:2012hy,Nitta:2012kk}
\beq
u_{\rm m-i-am} (x^1,z)
= (e^{- m (x^1 - x^1_1) + i \ph_1} 
 + e^{+ m (x^1 - x^1_2) + i \ph_2}) 
Z(z), \label{eq:monopole-instanton-monopole}
\eeq
with $Z(z)$ in Eq.~(\ref{eq:D-brane-soliton1}) 
for $k$ stretched instanton strings, 
or Eq.~(\ref{eq:D-brane-soliton2}) 
for $k_2$ stretched instanton strings 
and $k_1$ instanton strings attached from outside.
In the next section, we take $\ph_1 = \ph_2 +\pi$.

Before going to the next section, let us comment on 
$U(N)$ gauge theory. 
The Lagrangian (\ref{eq:Lagrangian}) can be extended to 
$U(N)$ gauge theory with $N \times N$ mass matrix, 
$M={\rm diag.}(m_1,m_2,\cdots,m_N)$ with $\sum_{A=1}^N m_A=0$. 
The vortex effective theory is the ${\bf C}P^{N-1}$ model 
with the twisted masses $M$.
One can construct $N-1$ monopole sheets parallel to each other 
as $N-1$ domain walls in the massive ${\bf C}P^{N-1}$ model \cite{Tong:2002hi,Isozumi:2004jc, Isozumi:2004va, Eto:2004vy},
with arbitrary number of instanton strings 
stretching between them or terminating on them 
\cite{Isozumi:2004vg}.
Such configurations were also considered in \cite{Tong:2005un} 
without putting them into a non-Abelian vortex.

\section{Pair annihilations of monopole and anti-monopole sheets
\label{sec:pair-annihilations}
}
\subsection{Instantons from annihilations of a monopole and an anti-monopole}

\begin{figure}
\begin{center}
\begin{tabular}{ccc}
\includegraphics[width=0.2\linewidth,keepaspectratio]{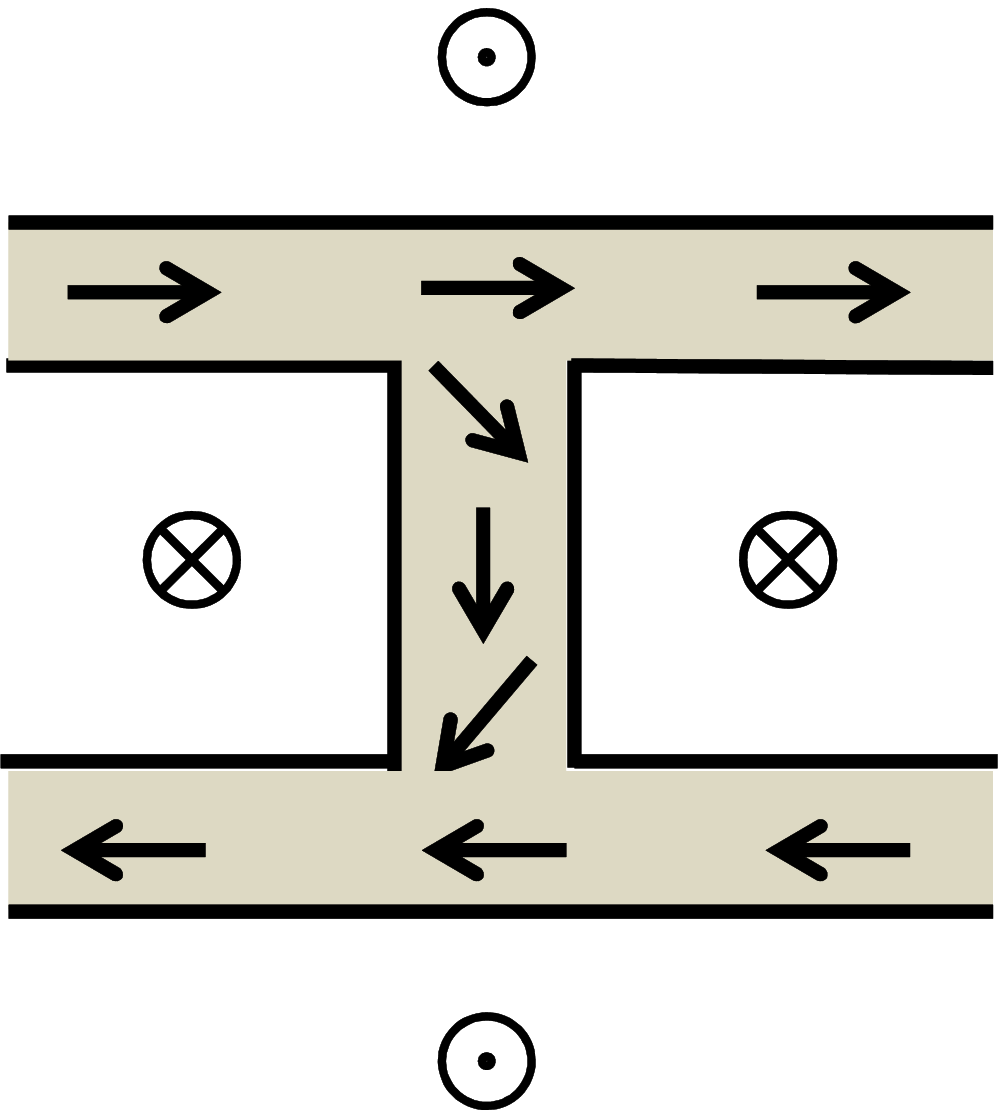}
&
\includegraphics[width=0.2\linewidth,keepaspectratio]{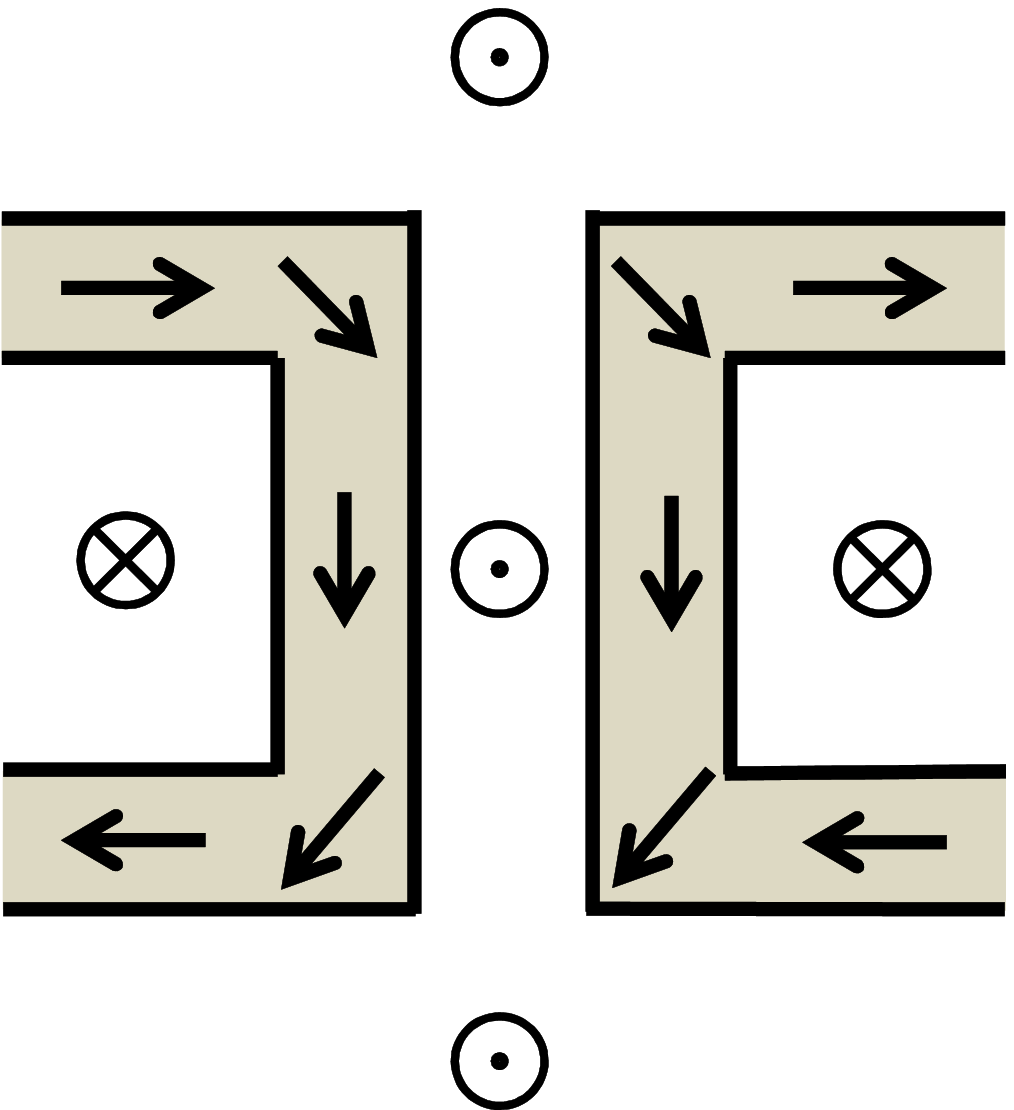}
&
\includegraphics[width=0.2\linewidth,keepaspectratio]{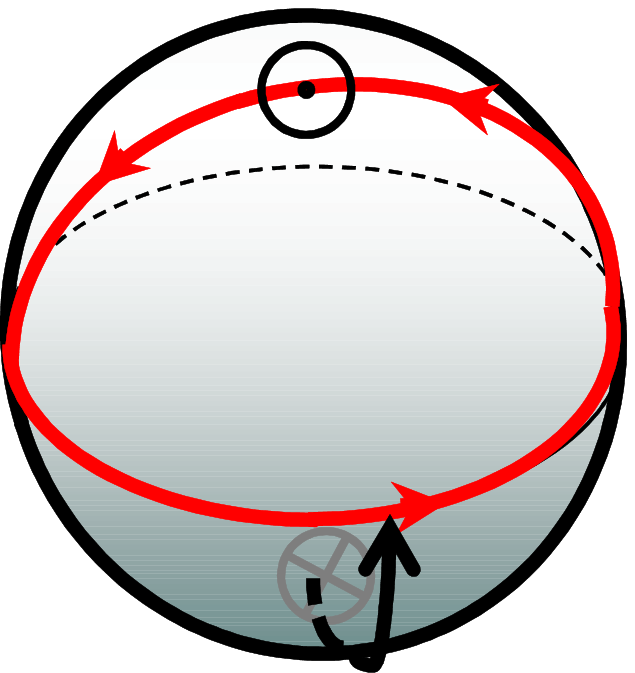}
\\
(a) & (c) & (e)\\
\includegraphics[width=0.2\linewidth,keepaspectratio]{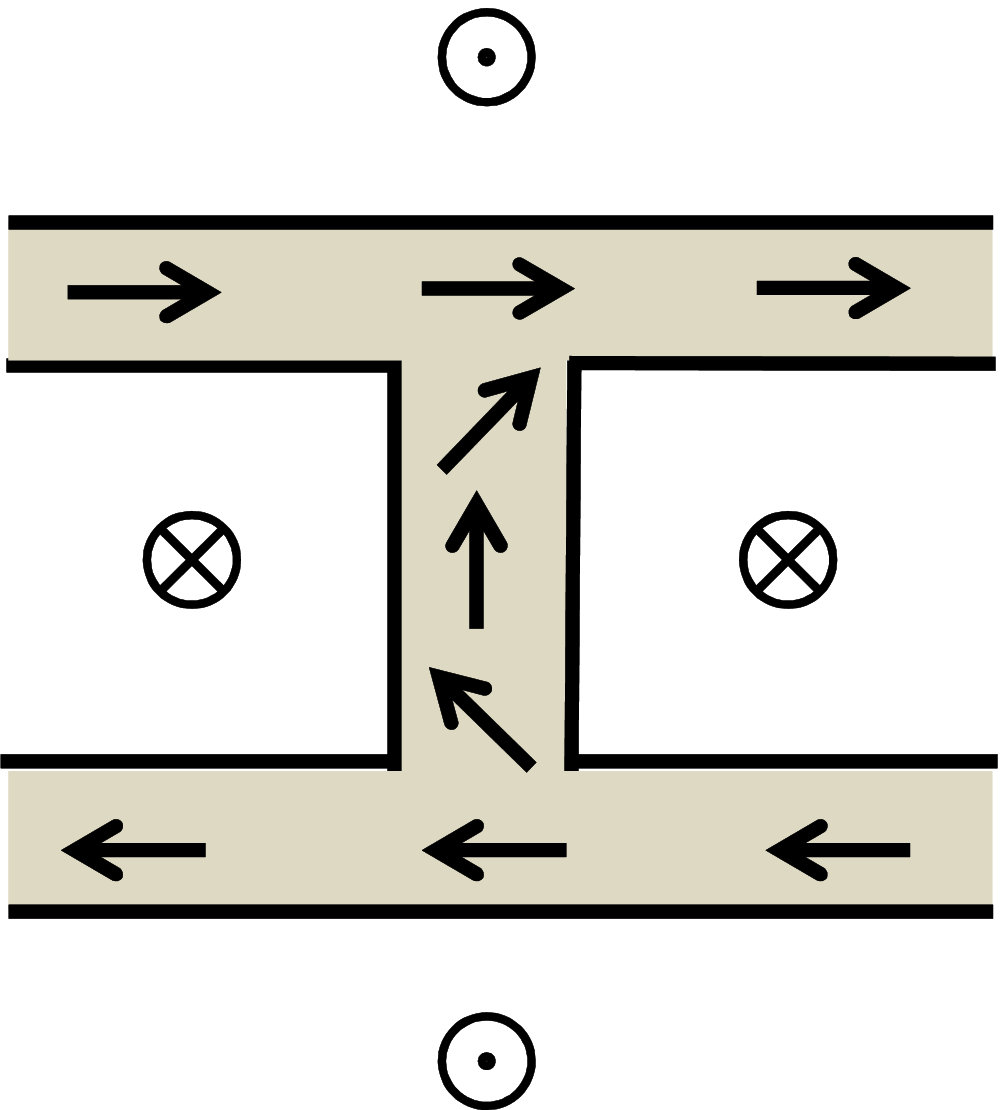}
&
\includegraphics[width=0.2\linewidth,keepaspectratio]{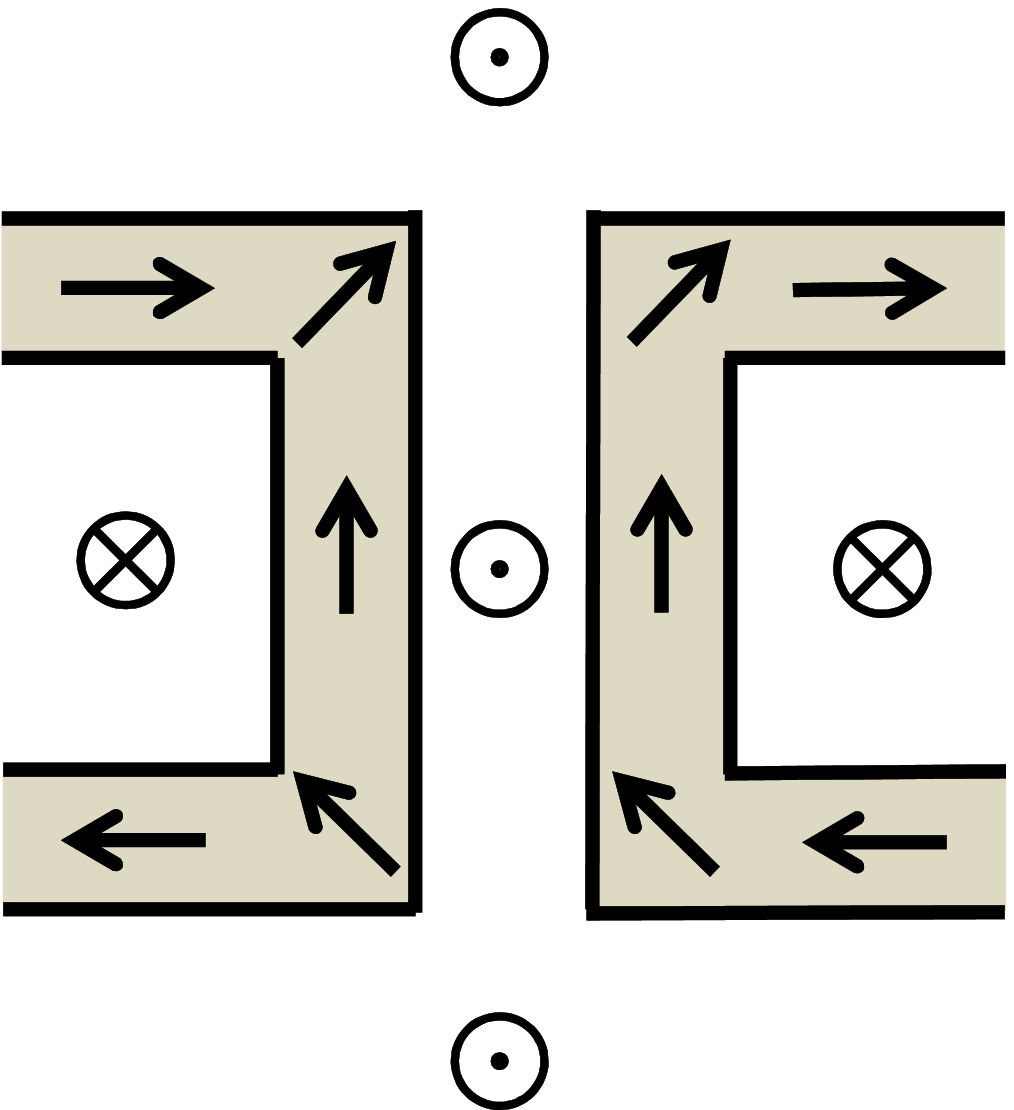}
&
\includegraphics[width=0.2\linewidth,keepaspectratio]{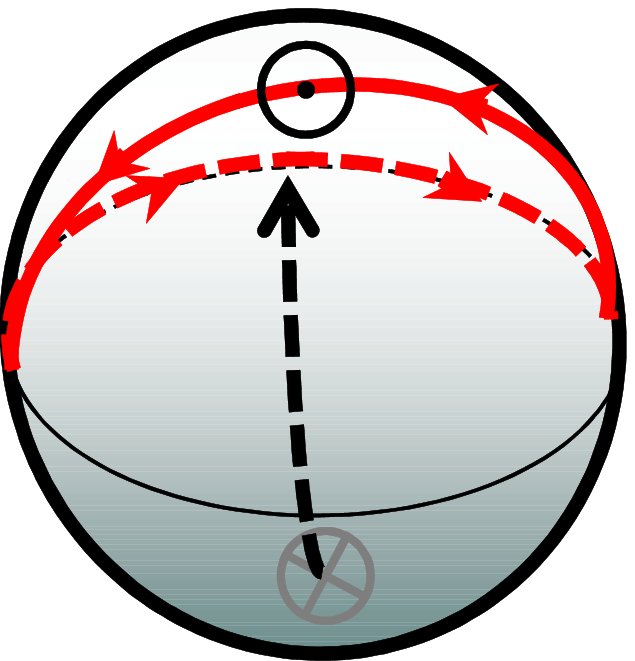}
\\
(b)& (d) & (f)
\end{tabular}
\caption{
Decaying processes of the monopole and anti-monopole.
(a), (b) A bridge is created between the monopole and the anti-monopole. 
In this process, there are two possibilities of 
the ${\bf C}P^1$ structure 
along the bridge. 
(c), (d) The upper and lower regions are connected 
by breaking the bridge. 
(e), (f) Accordingly, the loop in the ${\bf C}P^1$ target space 
is unwound in two ways.
\label{fig:wall-anti-wall-annihilation} 
} 
\end{center}
\end{figure}

The configuration of the monopole and anti-monopole 
is unstable to decay. 
We discuss the process of monopole decay following 
the corresponding case of domain walls \cite{Nitta:2012kj,Nitta:2012kk}. 
Here, we first discuss 
the case of $d=4+1$ dimensions in which  
monopoles and instantons are strings and particles, respectively. 
In the decaying process, the loop in the target space 
is unwound from the south pole $\otimes$ 
in Fig.~\ref{fig:brane-anti-brane-2d}(c), 
because we chose  the north pole $\odot$ as the vacuum state. 
The unwinding of the loop can be achieved in 
two topologically inequivalent processes, 
schematically shown in  
Fig.~\ref{fig:wall-anti-wall-annihilation}(e) and (f). 
In real space, 
at first, a bridge connecting the monopole and anti-monopole is created, 
as in Fig.~\ref{fig:wall-anti-wall-annihilation}(a) and (b).
Here, there exist two possibilities of the spin structure along the bridge, 
corresponding to the two inequivalent ways of 
the unwinding processes:
along the bridge in the $x^1$-direction, 
the spin rotates (a) clockwise or (b) anti-clockwise  on the equator 
of the ${\bf C}P^1$ target space. 
Subsequently, the bridge is broken into two pieces, 
as in Fig.~\ref{fig:wall-anti-wall-annihilation}(c) and (d),
between which the vacuum state $\odot$ is filled.
The two regions separated by the monopoles 
are connected through a hole created by the decay of the monopoles.
Once created, these holes grow with reducing the monopole energy.

Many holes are created everywhere in the monopole strings 
during the entire decaying process.
Let us focus a pair of two nearest holes.
\begin{figure}
\begin{center}
\includegraphics[width=0.7\linewidth,keepaspectratio]{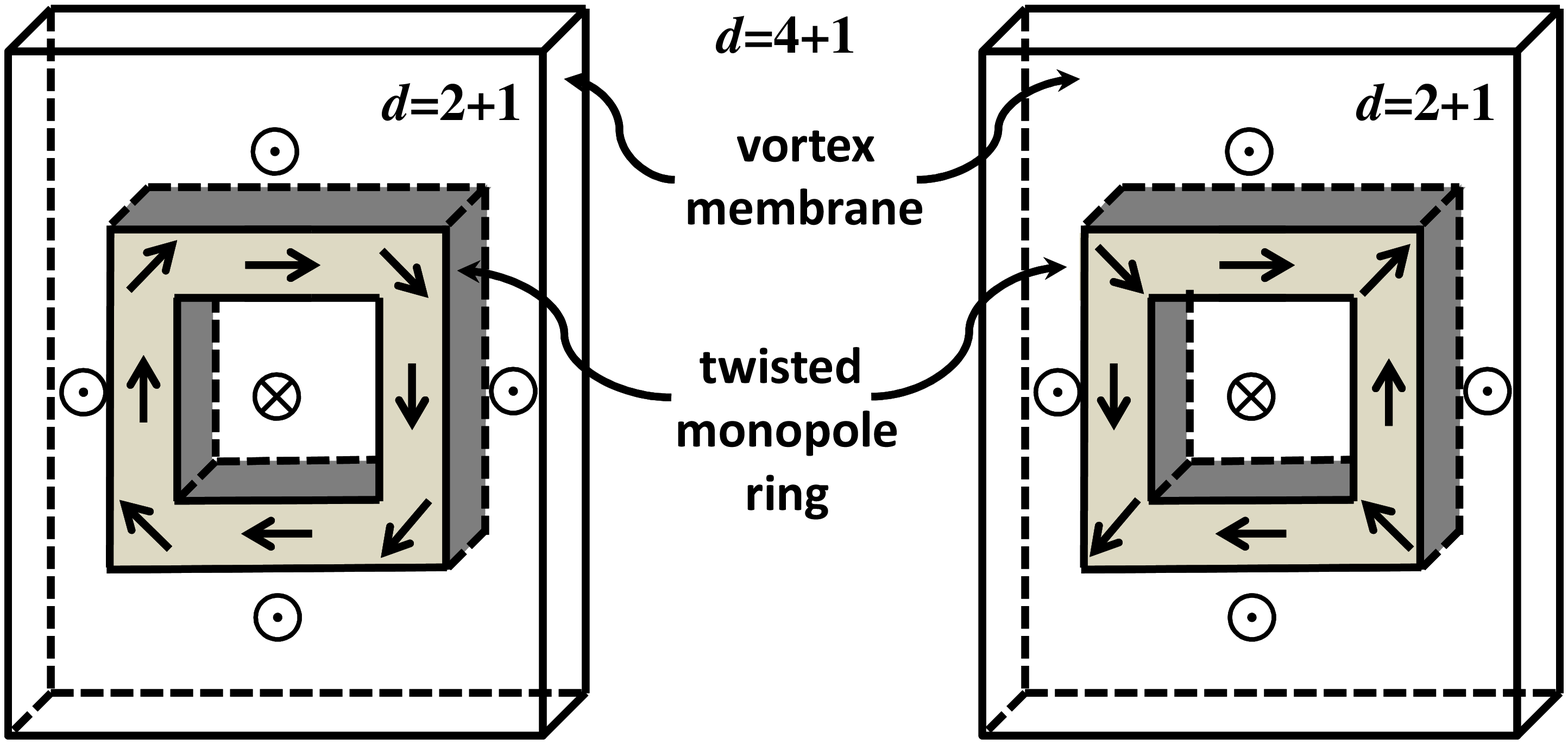}\\
(a) \hs{60} (b)
\caption{
(a) 
Twisted closed monopole strings inside a non-Abelian vortex in $d=4+1$. 
They are (a) an instanton and (b) an anti-instanton particles in $d=4+1$.
\label{fig:monopole-ring}}
\end{center}
\end{figure}
One can find that twisted closed monopole strings in Fig.~\ref{fig:monopole-ring} (a) and (b) are created; 
The configuration of Fig.~\ref{fig:monopole-ring} (a)  
is created by gluing 
Fig.~\ref{fig:wall-anti-wall-annihilation} (d) on the left 
and (c) on the right, 
while the one of Fig.~\ref{fig:monopole-ring} (b) is 
created by gluing 
Fig.~\ref{fig:wall-anti-wall-annihilation} (c) on the left 
and (d) on the right. 
Other two combinations give untwisted closed monopole strings, 
which are unstable to decay. 
The twisted closed monopole strings are nothing but (anti-)lumps in the ${\bf C}P^1$ model \cite{Polyakov:1975yp}
on the vortex world-volume, which 
can be shown to have a nontrivial winding 
in the second homotopy group $\pi_2 ({\bf C}P^1)\simeq {\bf Z}$; 
(a) and (b) belong to, respectively, 
$+1$ and $-1$ of $\pi_2 ({\bf C}P^1)$.
As already explained in Eq.~(\ref{eq:lumps}), 
the (anti-)lumps in the vortex effective theory 
correspond to (anti-)Yang-Mills instanton particles 
in the $d=4+1$ bulk \cite{Eto:2004rz,Fujimori:2008ee}.
Therefore we have seen that (anti-)instantons are created after 
the annihilation of the monopole and anti-monopole 
inside the vortex world-volume.

\begin{figure}[h]
\begin{center}
\begin{tabular}{cc}
\includegraphics[width=0.6\linewidth,keepaspectratio]{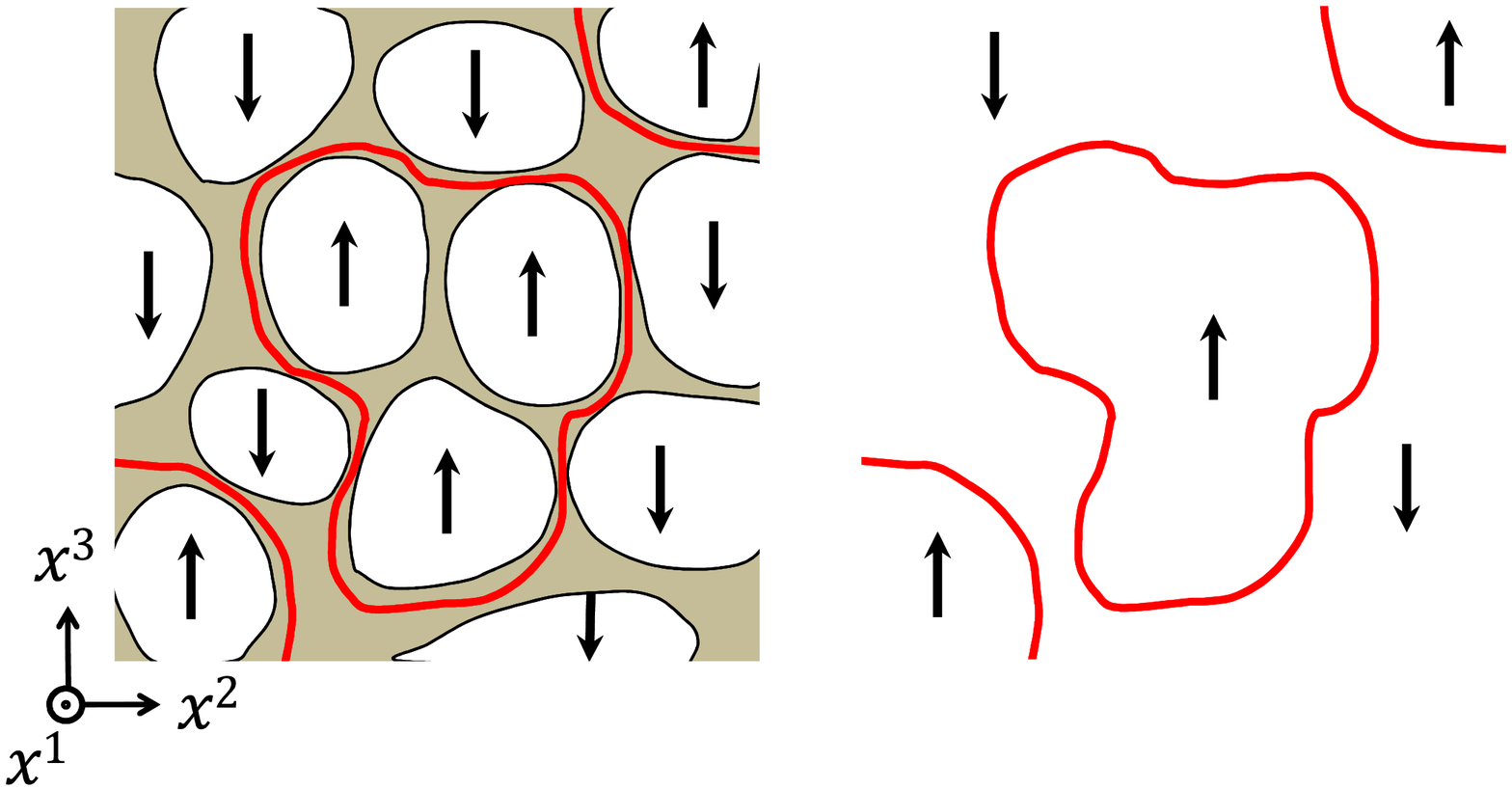} &
\includegraphics[width=0.4\linewidth,keepaspectratio]{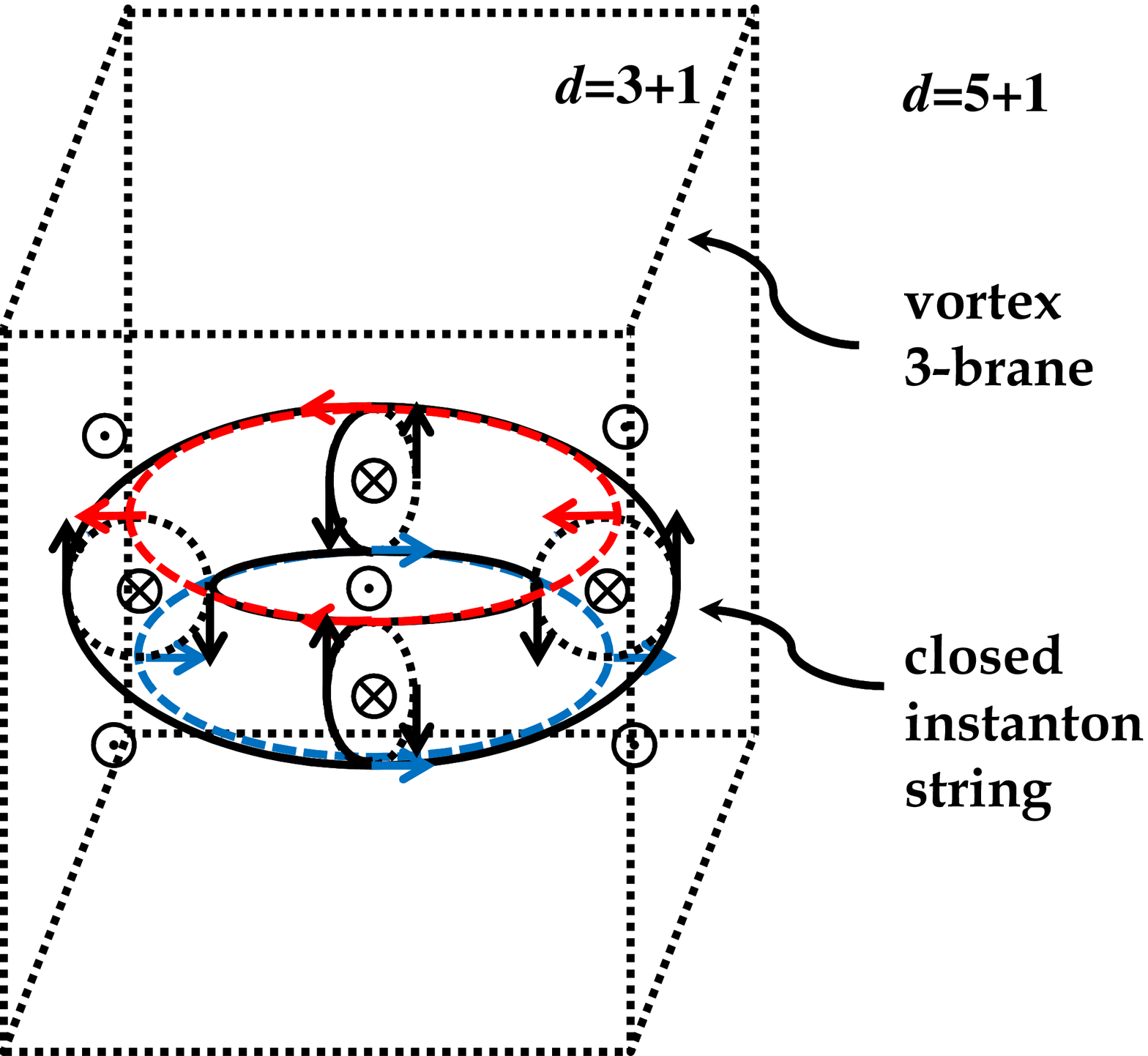}\\
(a) \hs{45} (b) & (c)\\
\end{tabular}
\caption{
Creation of a closed instanton string inside a non-Abelian vortex in $d=5+1$.
(a) Several two dimensional holes of two kinds labeled by 
$\uparrow$ and $\downarrow$, 
corresponding to 
Fig.~\ref{fig:wall-anti-wall-annihilation} (c) and (d), 
grow and meet. 
(b) A closed instanton loop remains on the edge of 
the $\uparrow$ and $\downarrow$ holes after sufficient growth of the holes. 
(c) A closed instanton loop inside a non-Abelian vortex in $d=5+1$.
\label{fig:monopole-ring-6d}
}
\end{center}
\end{figure}

In $d=5+1$ dimensions, 
monopoles are sheets 
and instantons are strings. 
When monopole and anti-monopole sheets 
annihilate in collision, 
two kinds of two-dimensional holes,  
corresponding to 
Fig.~\ref{fig:wall-anti-wall-annihilation} (c) and (d), 
are created in the monopole sheets, 
as in Fig.~\ref{fig:monopole-ring-6d} (a).  
When these holes grow sufficiently and they meet,  
there appear closed lump strings 
along the boundary of different kinds of holes, 
as in Fig.~\ref{fig:monopole-ring-6d} (b).
These closed lump strings in the ${\bf C}P^1$ vortex theory correspond to closed instanton strings 
in the $d=5+1$ dimensional bulk, see Fig \ref{fig:monopole-ring-6d}(c). 
The closed instanton strings are unstable in general 
and decay into elementary excitations in the end.

\subsection{Knotted instantons from annihilations of 
monopole and anti-monopole sheets with stretched instanton strings}

\begin{figure}[h]
\begin{center}
\begin{tabular}{cc}
\includegraphics[width=0.3\linewidth,keepaspectratio]{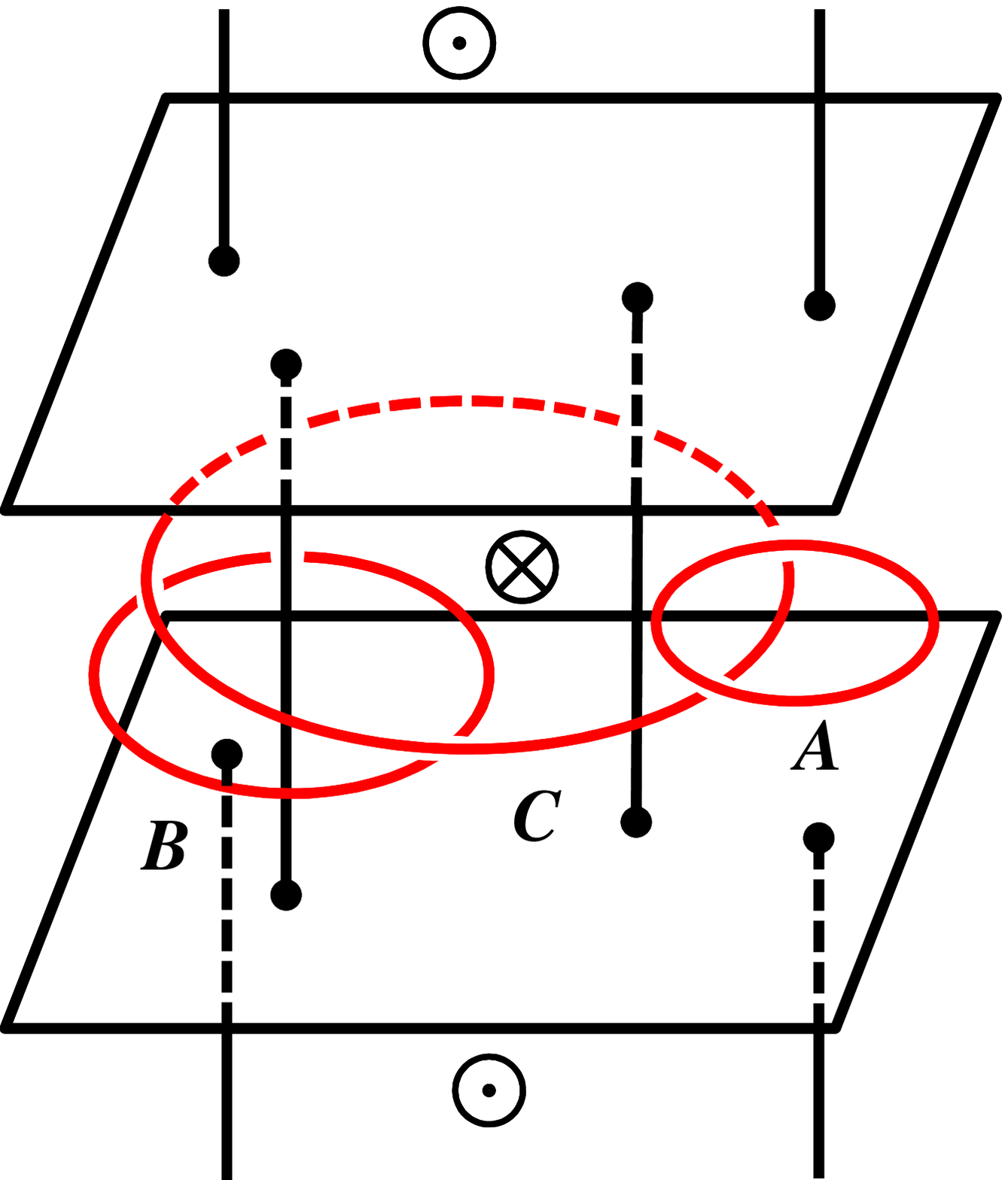}&
\hs{20}
\includegraphics[width=0.5\linewidth,keepaspectratio]{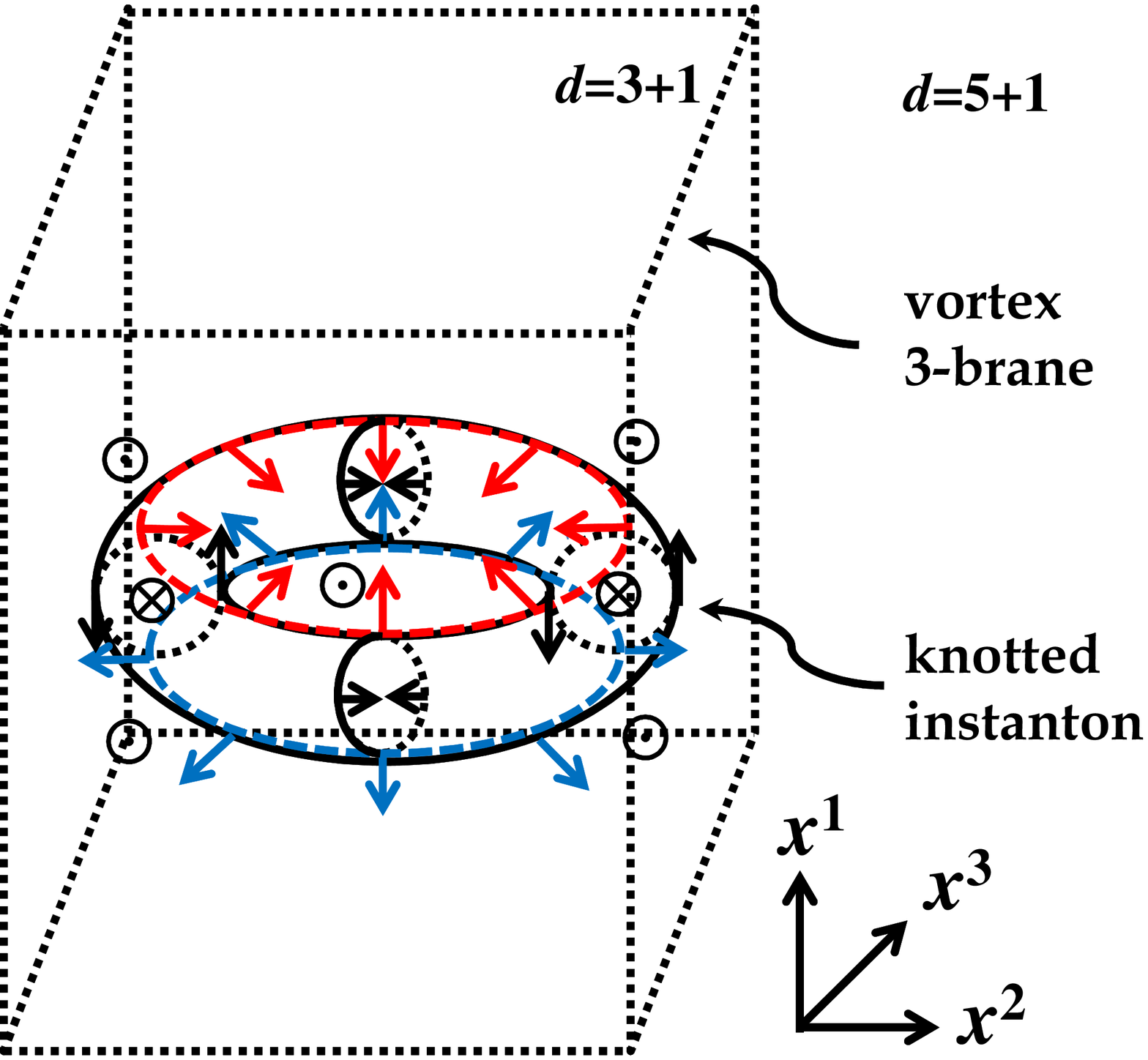}\\
(a) &(b)
\end{tabular}
\end{center}
\caption{(a) Loops in the monopole-instanton systems. 
While the loop A yields an untwisted closed instanton string 
in Fig.~\ref{fig:monopole-ring-6d}(c), 
the loop B (C) yields 
a closed instanton string twisted once (twice)
with the Hopf charge one (two).
(b) A closed instanton string twisted once, {\it i.e.}, knotted instanton.
\label{fig:monopole-pair-instanton+string}
} 
\end{figure}

In this section, 
we discuss the effect of instanton strings 
stretching between the monopole and anti-monopole sheets 
in Fig.~\ref{fig:brane-anti-brane-with-string2}. 
As in the case without a stretched instanton string,
the configuration itself is unstable to decay, 
and closed instanton strings are created. 
A created closed instanton string is not twisted,  
as before in Fig.~\ref{fig:monopole-ring-6d}(c),  
if it does not enclose
the stretched instanton strings, 
as the loop A in Fig.~\ref{fig:monopole-pair-instanton+string}(a).
However, if the closed instanton string encloses 
$n$ stretched instanton strings,  
as the loops B and C in Fig.~\ref{fig:monopole-pair-instanton+string}(a), 
it is twisted $n$ times.
A closed instanton string twisted once is shown 
in Fig.~\ref{fig:monopole-pair-instanton+string}(b). 
The vertical section of the torus in the $x^1$-$x^2$-plane  
is a pair of an instanton and an anti-instanton. 
Moreover, the presence of the stretched string 
implies that the phase modulus of the instanton 
winds anti-clockwise along the loops, 
as is indicated by the arrows on the top and bottom of the torus 
in Fig.~\ref{fig:monopole-pair-instanton+string}(b). 
When the instantons in the pair rotate along the $x^1$-axis, 
their phases are twisted and connected to each other 
at the $\pi$ rotation. 
Compare this with the untwisted closed instanton string in Fig.~\ref{fig:monopole-ring-6d}(c).

\begin{figure}[h]
\begin{center}
\includegraphics[width=0.3\linewidth,keepaspectratio]{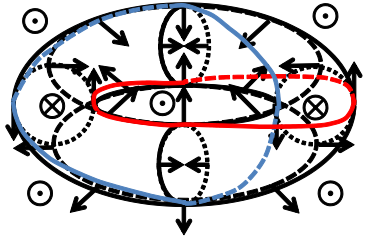}
\end{center}
\caption{\label{fig:linking}
The preimages of $\uparrow$ and $\downarrow$ make a link 
with  the linking number one, implying that the configuration is 
a knot soliton with the Hopf charge one, {\it i.e.}, the Hopfion.
} 
\end{figure}

This configuration is nothing but a Hopfion \cite{deVega:1977rk,Kundu:1982bc, Gladikowski:1996mb} 
in the ${\bf C}P^1$ model as the vortex effective theory.
One can confirm that the configuration has a unit Hopf charge as follows.
A preimage of a point on ${\bf C}P^1$ is a loop 
in real space.
When two loops of the preimages of two arbitrary points on ${\bf C}P^1$ 
have a linking number $n$, 
the configuration has a Hopf charge $n$.
See, e.g., Refs.~\cite{Kobayashi:2013bqa,Kobayashi:2013xoa, Kobayashi:2013aza} 
for a recent discussion. 
In Fig.~\ref{fig:linking}, we plot the preimages of 
$\uparrow$ and $\downarrow$, 
which are linked with a linking number one. 
Thus, we obtain a knot soliton with a Hopf charge of one 
(a Hopfion) inside the non-Abelian vortex.
Similarly, a created instanton string enclosing 
$n$ stretched instanton strings 
yields a knot soliton with a Hopf charge of $n$ 
inside the non-Abelian vortex.

To close this section, we discuss the stability of the knotted instanton.
Knot solitons in the Faddeev-Skyrme model \cite{Faddeev:1996zj} 
are stabilized by the Faddeev-Skyrme term containing quartic derivatives. 
The width of the lump strings (the size of the lumps) 
are fixed by the balance between 
the potential and quartic derivative terms, 
which make the solitons to shrink and to expand, respectively. 
The lump-string (lumps) whose size is fixed as this are known as 
baby-skyrme strings (baby skyrmions) 
\cite{Piette:1994ug,Piette:1994mh, Weidig:1998ii}, while the domain walls are essentially unchanged 
in the presence of the quartic derivative terms \cite{Kudryavtsev:1997nw,Harland:2007pb}.

The effective Lagrangian of the non-Abelian vortex 
with the quartic derivative correction 
can be written in genenal as \cite{Liu:2009rz}
\beq
{\cal L} =
 f^2 {\partial_{\mu} u^* \partial^{\mu} u - m^2 |u|^2 
  \over (1 + |u|^2)^2}  
+ c_1 { (\del_{\mu} u^* \del^{\mu} u)^2 - |\del_{\mu} u \del^{\mu} u|^2 
 \over (1+|u|^2)^4}  
+ c_2 { (\del_{\mu} u \del^{\mu} u^*)^2 
 \over (1+|u|^2)^4}  .
\eeq
By defining a three vector
\beq
{\bf n} = \Phi^\dagger {\bf \sigma} \Phi ,\quad
\Phi^T = (1,u)/\sqrt{1 + |u|^2}, \quad
{\bf n} \cdot {\bf n} = 1 ,
\eeq
the Lagrangian can be rewritten as
\beq
&& {\cal L} = \1{2}f^2
\del_{\mu}{\bf n}\cdot \del^{\mu} {\bf n} 
 - {\cal L}_4({\bf n})
 - V({\bf n}), \quad \\ 
\label{eq:Lagrangian}
&&{\cal L}_4 ({\bf n})
= c_1  \left[{\bf n} \cdot 
 (\partial_{\mu} {\bf n} \times \partial_{\nu} {\bf n} )\right]^2 
+ {c_2 \over 4} (\partial_{\mu} {\bf n} \cdot \partial^{\mu} {\bf n})^2
= c_1 (\partial_{\mu} {\bf n} \times \partial_{\nu} {\bf n} )^2 
+ {c_2 \over 4} (\partial_{\mu} {\bf n} \cdot \partial^{\mu} {\bf n})^2
, \\
&& V({\bf n})= m^2 (1-n_3^2) = m^2 (1-n_3)(1+n_3)
\eeq
in which we have used ${\bf n}\cdot \del {\bf n}=0$. 
The first term in ${\cal L}_4 ({\bf n})$ is nothing but 
the Faddeev-Skyrme term \cite{Faddeev:1996zj}
which stabilizes a knot soliton, 
while the second term, sometimes called the Gies term 
\cite{Gies:2001hk}, 
tends to distabilize a knot soliton \cite{Ferreira:2008nn,Ferreira:2009gj, Ferreira:2010jb}. 
The latter is inevitable in supersymmetric 
extension of the Faddeev-Skyrme model \cite{Bergshoeff:1984wb,Freyhult:2003zb,Adam:2011hj}.
In fact, the effective Lagrangian of the non-Abelian vortex 
with the quartic derivative correction 
was obtained recently in the case of the BPS vortex 
in supersymmetric theory \cite{Eto:2012qd}, in which 
it was found 
$c_2 = - 2c_1 = c\sim 0.830707 /g^2v^2$ and $f^2 = {4\pi/g^2}$.
In this case, the lump is scale invariant and has a size modulus, 
which implies that the knot soliton is not stabilized.

However we are working in a non-supersymmetric theory 
because the mass term in the Higgs field in forbidden in 
$d=5+1$ supersymmetric theory. 
By choosing parameters appropriately, 
there is a possibility that the vortex effective theory 
in a non-supersymmetric theory that contains 
a four derivative term stabilizing a knot soliton.
The construction of the vortex effective theory 
in non-supersymmetric theory up to a quartic term 
is technically challenging and  
remains as a future problem.

Finally, 
in $U(N)$ gauge theory, the vortex effective theory 
is the ${\bf C}P^{N-1}$ model with the twisted masses.
No topologically stable knot solitons exist because of 
$\pi_3 ({\bf C}P^{N-1})=0$. 

\section{Summary and discussion}\label{sec:summary}
In summary, when a pair of a monopole sheet and an anti-monopole sheet 
annihilates in collision inside a non-Abelian vortex,  
there appear closed instanton strings.
When an instanton string is stretched between the monopole sheets, 
a closed instanton string that encircles the stretched string 
becomes a twisted closed instanton string, which 
is a knotted instanton.

It was essential to create closed or knotted 
instanton strings that (anti-)monopoles 
have world-volumes in this dimensionality.
In $d=3+1$, (anti-)monopoles are particle-like, and 
therefore after pair annihilations they turn to just radiations.

In this paper, we have found that instanton strings in $d=5+1$ 
become knot solitons if twisted inside a non-Abelian vortex 
with $d=3+1$ dimensional world-volume. 
On the other hand, instanton particles in $d=4+1$ become Skyrmions inside a 
non-Abelian domain wall with $d=3+1$ dimensional world-volume 
\cite{Eto:2005cc}. 
These suggest a relation between knot solitons and Skyrmions 
since their holt solitons, a vortex and a domain wall,  
are related to each other by a duality, 
which is understood as the T-duality if embedded into string theory 
\cite{Eto:2006mz}.
If we take a T-duality along a non-Abelian vortex world-volume, 
instantons and monopole inside it are mapped to each other \cite{Eto:2004rz}, 
while we take a T-duality along a direction perpendicular 
to the vortex world-volume for the above duality. 

In this paper, we have worked in flat Minkowski space.
On the other hand, if one considers compactification to $S^1$ and/or $S^2$, some exact solutions are available.
In particular, 
the explicit monopole-antimonopole chain solutions on the space $S^1 \times S^2$ were constructed in Ref.~\cite{Popov:2008wi}. 
These exact solutions should be useful 
for understanding of decay of monopoles 
into instanton strings. 

Our knotted instantons can be regarded as higher dimensional 
generalization of the vortons.
The vorton is a twisted closed vortex string in $d=3+1$ 
\cite{Davis:1988jq,Radu:2008pp}. 
The vortons are usually expected to be stabilized by 
the Noether charge with giving a linear time dependence on 
the $U(1)$ Nambu-Goldstone mode of the vortex.
We may have to do the same for the stability of the knotted instanton, 
in which an instanton becomes a dyonic instanton \cite{Lambert:1999ua}.

Here we address what are not solved in this paper.

First, 
the topological charge $\pi_3 (S^2)$ of the knotted instanton 
can be defined in the vortex theory.  
However the topological charge in the bulk point of view is still unclear. 

Second, 
the dynamical stability of the knotted instanton should be clarified.
It remains as a future problem to derive 
the fourth derivative term stabilizing the knot soliton   
in the effective theory of the non-Abelian vortex. 
It is also an open question if the knotted instanton is stable
when the system is put into the unbroken phase 
in the limit of vanishing $v^2$, 
in which the host non-Abelian vortex disappears.

\section*{Acknowledgements}

I would like to thank Toshiaki Fujimori for a discussion.
This work is supported in part by 
Grant-in Aid for Scientific Research (No.~23740198) 
and by the ``Topological Quantum Phenomena'' 
Grant-in Aid for Scientific Research 
on Innovative Areas (No.~23103515)  
from the Ministry of Education, Culture, Sports, Science and Technology 
(MEXT) of Japan. 



\begin{thebibliography}{99}

\bibitem{'tHooft:1974qc}
  G.~'t Hooft,
  Nucl.\ Phys.\  B {\bf 79}, 276 (1974).

\bibitem{Polyakov:1974ek}
  A.~M.~Polyakov,
  JETP Lett.\  {\bf 20}, 194 (1974)
  [Pisma Zh.\ Eksp.\ Teor.\ Fiz.\  {\bf 20}, 430 (1974)].

\bibitem{Belavin:1975fg}
  A.~A.~Belavin, A.~M.~Polyakov, A.~S.~Shvarts and Yu.~S.~Tyupkin,
  Phys.\ Lett.\  B {\bf 59}, 85 (1975).

\bibitem{Witten:1995gx} 
  E.~Witten,
Nucl.\ Phys.\ B {\bf 460}, 541 (1996)  [hep-th/9511030].  

\bibitem{Douglas:1995bn} 
  M.~R.~Douglas,
In *Cargese 1997, Strings, branes and dualities* 267-275  [hep-th/9512077].  

\bibitem{Green:1996qg} 
  M.~B.~Green and M.~Gutperle,
Phys.\ Lett.\ B {\bf 377}, 28 (1996)  [hep-th/9602077]. 
\bibitem{Diaconescu:1996rk} 
  D.~-E.~Diaconescu,
Nucl.\ Phys.\ B {\bf 503}, 220 (1997)  [hep-th/9608163].  

\bibitem{Preskill:1992ck} 
  J.~Preskill and A.~Vilenkin,
Phys.\ Rev.\ D {\bf 47}, 2324 (1993). 
\bibitem{Vilenkin:1982hm} 
  A.~Vilenkin,
Nucl.\ Phys.\ B {\bf 196}, 240 (1982).  

\bibitem{Sen:2004nf} 
  A.~Sen,
Int.\ J.\ Mod.\ Phys.\ A {\bf 20}, 5513 (2005)
  [hep-th/0410103].  

\bibitem{Nitta:2012kj} 
  M.~Nitta,
Phys.\ Rev.\ D {\bf 85}, 101702 (2012)  [arXiv:1205.2442 [hep-th]].  

\bibitem{Takeuchi:2012ee} 
  H.~Takeuchi, K.~Kasamatsu, M.~Tsubota and M.~Nitta,
Phys.\ Rev.\ Lett.\  {\bf 109}, 245301 (2012)  [arXiv:1205.2330 [cond-mat.quant-gas]].  

\bibitem{Takeuchi:2012ec} 
  H.~Takeuchi, K.~Kasamatsu, M.~Nitta and M.~Tsubota,
J.\ Low.\ Temp.\ Phys.\  {\bf 162}, 243 (2011) 
  [arXiv:1205.2328 [cond-mat.quant-gas]].  


\bibitem{Hanany:2003hp}
  A.~Hanany and D.~Tong,
  JHEP {\bf 0307}, 037 (2003)
  [arXiv:hep-th/0306150].
\bibitem{Auzzi:2003fs}
  R.~Auzzi, S.~Bolognesi, J.~Evslin, K.~Konishi and A.~Yung,
  Nucl.\ Phys.\  B {\bf 673}, 187 (2003)
  [arXiv:hep-th/0307287].

\bibitem{Eto:2005yh} 
  M.~Eto, Y.~Isozumi, M.~Nitta, K.~Ohashi and N.~Sakai,
Phys.\ Rev.\ Lett.\  {\bf 96}, 161601 (2006)
  [hep-th/0511088]. 
\bibitem{Eto:2006cx} 
  M.~Eto, K.~Konishi, G.~Marmorini, M.~Nitta, K.~Ohashi, W.~Vinci and N.~Yokoi,
Phys.\ Rev.\ D {\bf 74}, 065021 (2006)  [hep-th/0607070].  

\bibitem{Tong:2005un} 
  D.~Tong,
hep-th/0509216.  

\bibitem{Eto:2006pg} 
  M.~Eto, Y.~Isozumi, M.~Nitta, K.~Ohashi and N.~Sakai,
J.\ Phys.\ A {\bf 39}, R315 (2006)  [hep-th/0602170].  

\bibitem{Shifman:2007ce} 
  M.~Shifman and A.~Yung,
Rev.\ Mod.\ Phys.\  {\bf 79}, 1139 (2007)  [hep-th/0703267].  
\bibitem{Shifman:2009zz} 
  M.~Shifman and A.~Yung,
  ``Supersymmetric solitons,''  Cambridge, UK: Cambridge Univ. Pr. (2009) 259 p.

\bibitem{Eto:2004rz} 
  M.~Eto, Y.~Isozumi, M.~Nitta, K.~Ohashi and N.~Sakai,
Phys.\ Rev.\ D {\bf 72}, 025011 (2005)
 [hep-th/0412048].  

\bibitem{Fujimori:2008ee} 
  T.~Fujimori, M.~Nitta, K.~Ohta, N.~Sakai and M.~Yamazaki,
Phys.\ Rev.\ D {\bf 78}, 105004 (2008)  [arXiv:0805.1194 [hep-th]].  

\bibitem{Tong:2003pz} 
  D.~Tong,
Phys.\ Rev.\ D {\bf 69}, 065003 (2004)
  [hep-th/0307302].  

\bibitem{Polyakov:1975yp} 
  A.~M.~Polyakov and A.~A.~Belavin,
JETP Lett.\  {\bf 22}, 245 (1975)  [Pisma Zh.\ Eksp.\ Teor.\ Fiz.\  {\bf 22}, 503 (1975)].  

\bibitem{Abraham:1992vb} 
  E.~R.~C.~Abraham and P.~K.~Townsend,
Phys.\ Lett.\ B {\bf 291}, 85 (1992).  
\bibitem{Abraham:1992qv} 
  E.~R.~C.~Abraham and P.~K.~Townsend,
Phys.\ Lett.\ B {\bf 295}, 225 (1992).  

\bibitem{Arai:2002xa} 
  M.~Arai, M.~Naganuma, M.~Nitta and N.~Sakai,
 Nucl.\ Phys.\ B {\bf 652}, 35 (2003)  [hep-th/0211103].  
\bibitem{Arai:2003es} 
  M.~Arai, M.~Naganuma, M.~Nitta and N.~Sakai,
 In *Arai, A. (ed.) et al.: A garden of quanta* 299-325  [hep-th/0302028].  

\bibitem{Faddeev:1996zj}
  L.~D.~Faddeev and A.~J.~Niemi,
  Nature {\bf 387}, 58 (1997)
  [arXiv:hep-th/9610193].

\bibitem{Radu:2008pp} 
  E.~Radu and M.~S.~Volkov,
Phys.\ Rept.\  {\bf 468}, 101 (2008)  [arXiv:0804.1357 [hep-th]].  

\bibitem{Nitta:2012kk} 
  M.~Nitta,
Phys.\ Rev.\ D {\bf 85}, 121701 (2012)  [arXiv:1205.2443 [hep-th]].  

\bibitem{Davis:1988jq} 
  R.~L.~Davis and E.~P.~S.~Shellard,
Phys.\ Lett.\ B {\bf 209}, 485 (1988).  

\bibitem{Nitta:2012hy} 
  M.~Nitta, K.~Kasamatsu, M.~Tsubota and H.~Takeuchi,
Phys.\ Rev.\ A {\bf 85}, 053639 (2012)  [arXiv:1203.4896 [cond-mat.quant-gas]].  

\bibitem{Eto:2006uw} 
  M.~Eto, Y.~Isozumi, M.~Nitta, K.~Ohashi and N.~Sakai,
Phys.\ Rev.\ D {\bf 73}, 125008 (2006)  [hep-th/0602289].  

\bibitem{Gauntlett:2000de} 
  J.~P.~Gauntlett, R.~Portugues, D.~Tong and P.~K.~Townsend,
Phys.\ Rev.\ D {\bf 63}, 085002 (2001)
  [hep-th/0008221].  

\bibitem{Shifman:2002jm} 
  M.~Shifman and A.~Yung,
Phys.\ Rev.\ D {\bf 67}, 125007 (2003)  [hep-th/0212293].  

\bibitem{Isozumi:2004vg} 
  Y.~Isozumi, M.~Nitta, K.~Ohashi and N.~Sakai,
Phys.\ Rev.\ D {\bf 71}, 065018 (2005)
  [hep-th/0405129].  

\bibitem{Eto:2004zc} 
  M.~Eto, N.~Maru and N.~Sakai,
Nucl.\ Phys.\ B {\bf 696}, 3 (2004)  [hep-th/0404114].  

\bibitem{Callan:1997kz} 
  C.~G.~Callan and J.~M.~Maldacena,
 Nucl.\ Phys.\ B {\bf 513}, 198 (1998)
  [hep-th/9708147].  
\bibitem{Gibbons:1997xz} 
  G.~W.~Gibbons,
Nucl.\ Phys.\ B {\bf 514}, 603 (1998)
  [hep-th/9709027].  
\bibitem{Hashimoto:1997px} 
  A.~Hashimoto,
Phys.\ Rev.\ D {\bf 57}, 6441 (1998).



\bibitem{Tong:2002hi} 
  D.~Tong,
Phys.\ Rev.\ D {\bf 66}, 025013 (2002)  [hep-th/0202012].  
\bibitem{Isozumi:2004jc} 
  Y.~Isozumi, M.~Nitta, K.~Ohashi and N.~Sakai,
Phys.\ Rev.\ Lett.\  {\bf 93}, 161601 (2004)  [hep-th/0404198].  
\bibitem{Isozumi:2004va} 
  Y.~Isozumi, M.~Nitta, K.~Ohashi and N.~Sakai,
Phys.\ Rev.\ D {\bf 70}, 125014 (2004)  [hep-th/0405194].  
\bibitem{Eto:2004vy} 
  M.~Eto, Y.~Isozumi, M.~Nitta, K.~Ohashi, K.~Ohta and N.~Sakai,
Phys.\ Rev.\ D {\bf 71}, 125006 (2005)  [hep-th/0412024].  

\bibitem{deVega:1977rk} 
  H.~J.~de Vega,
Phys.\ Rev.\ D {\bf 18}, 2945 (1978).  
\bibitem{Kundu:1982bc} 
  A.~Kundu and Y.~P.~Rybakov,
J.\ Phys.\ A A {\bf 15}, 269 (1982).  
\bibitem{Gladikowski:1996mb} 
  J.~Gladikowski and M.~Hellmund,
Phys.\ Rev.\ D {\bf 56}, 5194 (1997).  

\bibitem{Kobayashi:2013bqa} 
  M.~Kobayashi and M.~Nitta,
  ``Toroidal domain walls as Hopfions,''
  arXiv:1304.4737 [hep-th].
\bibitem{Kobayashi:2013xoa} 
  M.~Kobayashi and M.~Nitta,
  ``Torus knots as Hopfions,''
  Phys.\ Lett.\ B (to appear) [arXiv:1304.6021 [hep-th]. 
\bibitem{Kobayashi:2013aza} 
  M.~Kobayashi and M.~Nitta,
  ``Winding Hopfions on R$^{2} \times S^{1}$,''
  Nucl.\ Phys.\ B {\bf 876}, 605 (2013)
  [arXiv:1305.7417 [hep-th]].


\bibitem{Piette:1994ug}
  B.~M.~A.~Piette, B.~J.~Schroers and W.~J.~Zakrzewski,
  Z.\ Phys.\  C {\bf 65}, 165 (1995).
\bibitem{Piette:1994mh}
  B.~M.~A.~Piette, B.~J.~Schroers and W.~J.~Zakrzewski,
  Nucl.\ Phys.\  B {\bf 439}, 205 (1995).
  [arXiv:hep-ph/9410256].
\bibitem{Weidig:1998ii}
  T.~Weidig,
Nonlinearity {\bf 12}, 1489-1503 (1999)
  [arXiv:hep-th/9811238].

\bibitem{Kudryavtsev:1997nw}
  A.~E.~Kudryavtsev, B.~M.~A.~Piette and W.~J.~Zakrzewski,
  Nonlinearity {\bf 11}, 783 (1998)
  [arXiv:hep-th/9709187].
\bibitem{Harland:2007pb}
  D.~Harland and R.~S.~Ward,
  Phys.\ Rev.\  D {\bf 77}, 045009 (2008)
  [arXiv:0711.3166 [hep-th]].

\bibitem{Liu:2009rz} 
  L.~-X.~Liu and M.~Nitta,
Int.\ J.\ Mod.\ Phys.\ A {\bf 27}, 1250097 (2012)  [arXiv:0912.1292 [hep-th]].  


\bibitem{Gies:2001hk}
  H.~Gies,
  Phys.\ Rev.\  D {\bf 63}, 125023 (2001)
  [arXiv:hep-th/0102026].


\bibitem{Ferreira:2008nn}
  L.~A.~Ferreira,
  JHEP {\bf 0905}, 001 (2009) [arXiv:0809.4303 [hep-th]].
\bibitem{Ferreira:2009gj}
  L.~A.~Ferreira, N.~Sawado, K.~Toda,
  JHEP {\bf 0911}, 124 (2009)
  [arXiv:0908.3672 [hep-th]].
\bibitem{Ferreira:2010jb}
  L.~A.~Ferreira, P.~Klimas,
  JHEP {\bf 1010}, 008 (2010)
  [arXiv:1007.1667 [hep-th]].

\bibitem{Bergshoeff:1984wb} 
  E.~A.~Bergshoeff, R.~I.~Nepomechie and H.~J.~Schnitzer,
Nucl.\ Phys.\ B {\bf 249}, 93 (1985).  
\bibitem{Freyhult:2003zb}
  L.~Freyhult,
  Nucl.\ Phys.\  B {\bf 681}, 65 (2004) 
  [arXiv:hep-th/0310261].

\bibitem{Adam:2011hj} 
In the case of lower dimensional Faddeev-Skyrme model,
known as the baby Skyrme model, 
such extra term is not needed for less supersymmety in $d=2+1$, 
that is two supercharges:
  C.~Adam, J.~M.~Queiruga, J.~Sanchez-Guillen and A.~Wereszczynski,
Phys.\ Rev.\ D {\bf 84}, 025008 (2011)
  [arXiv:1105.1168 [hep-th]].  

\bibitem{Eto:2012qd} 
  M.~Eto, T.~Fujimori, M.~Nitta, K.~Ohashi and N.~Sakai,
Prog.\ Theor.\ Phys.\  {\bf 128}, 67 (2012)  [arXiv:1204.0773 [hep-th]].  


\bibitem{Eto:2005cc} 
  M.~Eto, M.~Nitta, K.~Ohashi and D.~Tong,
Phys.\ Rev.\ Lett.\  {\bf 95}, 252003 (2005)  [hep-th/0508130].  

\bibitem{Eto:2006mz} 
  M.~Eto, T.~Fujimori, Y.~Isozumi, M.~Nitta, K.~Ohashi, K.~Ohta and N.~Sakai,
Phys.\ Rev.\ D {\bf 73}, 085008 (2006)  [hep-th/0601181].  

\bibitem{Popov:2008wi} 
  A.~D.~Popov,
  Phys.\ Rev.\ D {\bf 77}, 125026 (2008)
  [arXiv:0803.3320 [hep-th]].

\bibitem{Lambert:1999ua} 
  N.~D.~Lambert and D.~Tong,
Phys.\ Lett.\ B {\bf 462}, 89 (1999)  [hep-th/9907014].  

\end{thebibliography}
\end{document}